\documentclass[nature
,showpacs
,superscriptaddress
]
{revtex4-2}
\usepackage{graphicx,subfigure,bbm}

\usepackage{amsmath}
\usepackage{amsthm}
\usepackage{amssymb}
\usepackage{array,bm}
\usepackage{amsfonts} %
\usepackage{graphicx}

\usepackage{caption}
\usepackage{ragged2e}

\DeclareCaptionLabelFormat{myformat}{\textbf{FIG. #2}}

\usepackage[colorlinks,
            linkcolor=black,
            citecolor=black,
            urlcolor=blue
            ]{hyperref}
\def\figureautorefname{FIG.}

\usepackage{cleveref}
\crefname{equation}{Eq.}{Eqs.}

\usepackage{fancyhdr}
\usepackage{float}
\usepackage{color}
\usepackage{extarrows}
\usepackage{enumerate}
\usepackage{epstopdf}
\usepackage{booktabs}
\usepackage{array}

\usepackage{natbib}

\def\(({\left(}
\def\)){\right)}
\def\[[{\left[}
\def\]]{\right]}

\newcommand{\be}{\begin{equation}}
\newcommand{\ee}{\end{equation}}
\newcommand{\bea}{\begin{eqnarray}}
\newcommand{\eea}{\end{eqnarray}}

\newcommand{\supplementarysection}{%
  \setcounter{figure}{0}
  \let\oldthefigure\thefigure
\renewcommand{\figurename}{Supplement FIG.}
\setcounter{table}{0}
\renewcommand{\tablename}{Supplementary Table}
  \setcounter{equation}{0}
  \let\oldtheequation\theequation
  \renewcommand{\theequation}{S\oldtheequation}
  \section{Supplementary Information}
 \def\figureautorefname{Supplement FIG.}
 \def\tableautorefname{Supplementary Table}
}

\DeclareMathAlphabet{\varmathbb}{U}{bbold}{m}{n}

\begin{document}

\title{Learning noise-induced transitions by multi-scaling reservoir computing}

\author{Zequn Lin}
\email[These authors contributed equally]{}
\affiliation{Faculty of Arts and Sciences, Beijing Normal University, Zhuhai 519087, China}

\author{Zhaofan Lu}
\email[These authors contributed equally]{}
\affiliation{Faculty of Arts and Sciences, Beijing Normal University, Zhuhai 519087, China}

\author{Zengru Di}
\affiliation{International Academic Center of Complex Systems, Beijing Normal University, Zhuhai 519087, China}

\author{Ying Tang}
\email{jamestang23@gmail.com}
\affiliation{International Academic Center of Complex Systems, Beijing Normal University, Zhuhai 519087, China}

\begin{abstract}
Noise is usually regarded as adversarial to extract the effective dynamics from time series, such that the conventional data-driven approaches usually aim at learning the dynamics by mitigating the noisy effect. However, noise can have a functional role of driving transitions between stable states underlying many natural and engineered stochastic dynamics. To capture such stochastic transitions from data, we find that leveraging a machine learning model, reservoir computing as a type of recurrent neural network, can learn noise-induced transitions. We develop a concise training protocol for tuning hyperparameters, with a focus on a pivotal hyperparameter controlling the time scale of the reservoir dynamics. The trained model generates accurate statistics of transition time and the number of transitions. The approach is applicable to a wide class of systems, including a bistable system under a double-well potential, with either white noise or colored noise. It is also aware of the asymmetry of the double-well potential, the rotational dynamics caused by non-detailed balance, and transitions in multi-stable systems. For the experimental data of protein folding, it learns the transition time between folded states, providing a possibility of predicting transition statistics from a small dataset. The results demonstrate the capability of machine-learning methods in capturing noise-induced phenomena. 
\end{abstract}

\maketitle

\section{Introduction}

Noise-induced transitions are ubiquitous in nature and occur in diverse systems with multi-stable states \cite{horsthemke1984noise}. Examples include switches between different voltage and current states in the circuit \cite{semenov2016noise},  noisy genetic switches \cite{PhysRevLett.106.248102}, noise-induced biological homochirality of early life self-replicators \cite{PhysRevLett.115.158101}, protein conformational transitions \cite{qian2002discrete,tapia2023enhanced}, and chemical reactions \cite{RevModPhys.62.251} with the multi-stable probability distribution \cite{tang2023neural}. Learning noise-induced transitions is vital for understanding critical phenomena of these systems. In many scenarios, only time series are available without mathematical equations known in prior. 
To effectively learn and predict noise-induced transitions from time series, there is also a challenge of discerning dynamics with both slow and fast time scales: fast relaxation around distinct stable states and slow transitions between them, where the fast time-scale signals are often referred to noise \cite{forgoston2018primer, hartmann2013characterization}. Consequently, it remains elusive to learn stochastic transitions from time series in general.

Recently, many efforts have been made to learn the dynamics from data by machine-learning methods \cite{tanaka2019recent,xiong2019chaotic,karniadakis2021physics,zhao2021inferring,li2023meta}. One type of approach uses the automatic differentiation for identifying nonlinear dynamics, denoising time-series data, and parameterizing the noisy probability distribution from data \cite{kaheman2022automatic}. Due to the non-convexity of the optimization problem, the method may struggle to robustly handle large function libraries for the regression. Another type of approach employs physics-informed neural networks for data-driven solutions and discoveries of partial differential equations  \cite{raissi2019physics,cuomo2022scientific}, or Koopman eigenfunctions from data \cite{lusch2018deep}.
However, the method requires an extensive quantity of data to train the deep neural network, alongside precise adjustment and refinement of the network. Despite the broad application of the aforementioned methods, to our knowledge, they have not been utilized in studying noise-induced transitions.

To investigate whether machine-learning methods can capture and predict noise-induced transitions, we start with one machine-learning architecture, reservoir computing (RC) \cite{jaeger2001echo,maass2002real}. The training of reservoir computer is a simple linear regression, which is less computationally expensive than the neural network that requires the back propagation. The input layer of the reservoir transforms time series into the space of the reservoir network, while the output layer transforms the variables of the reservoir back to time series. The output layer is trained to minimize the difference between the input and output, by tuning the hyperparameters. The reservoir computing is particularly effective for learning dynamical systems \cite{pathak2018model,kim2023neural}, including chaotic systems \cite{jaeger2004harnessing,zimmermann2018observing,fan2020long,kim2021teaching}. A recent research started to apply the reservoir computing to stochastic resonance \cite{zhai2023emergence}, however, the functional role of noise in shifting dynamics between stable states has not been investigated. There is one previous attempt on employing reservoir computing for noise-induced transitions \cite{Lim2020Predicting}. Nevertheless, it relies on an impractical assumption on knowing the equation for the deterministic part of dynamics, and then employs reservoir computer to learn the separated fast time-scale series. 
In practice, one usually lacks a prior knowledge about the deterministic dynamics, and in some cases, this part even cannot be directly described by an equation \cite{tapia2023enhanced}. Thus, the question remains that can we forecast noise-induced transitions solely based on data without any prior knowledge of the underlying equation?

In this paper, we develop a framework of multi-scaling reservoir computing for learning noise-induced transitions in a model-free manner. Our method is inspired by the capability of reservoir computer to model dynamical systems \cite{jaeger2001echo,pathak2018model,carroll2018using,nakai2018machine,weng2019synchronization,maass2002real,grigoryeva2018echo,zhang2021learning}, and especially the hyperparameter $\alpha$ in the reservoir was found to determine the time scale of reservoir dynamics \cite{tanaka2021reservoir}. Given a multi-scale time series, we can thus tune the hyperparameter $\alpha$ to match the slowly time-scale dynamics. After the reservoir captures the slowly time-scale dynamics by fitting the output layer matrix, we can separate the fast time-scale series as a noise distribution. During the predicting phase, we utilize the trained reservoir computer to simulate the slowly time-scale dynamics, and then add back the noise sampled from the separated noise distribution (for white noise) or learnt from the second reservoir (for colored noise). The whole protocol is iterated over time points as the rolling prediction. Notably, the present method is different from previous work that regards noise merely as a disturbance \cite{kaheman2022automatic,du2023inferring}, and instead focuses on capturing noise-induced transitions from the data.

\begin{figure}[htbp]
    \centering  \includegraphics[width=\textwidth]{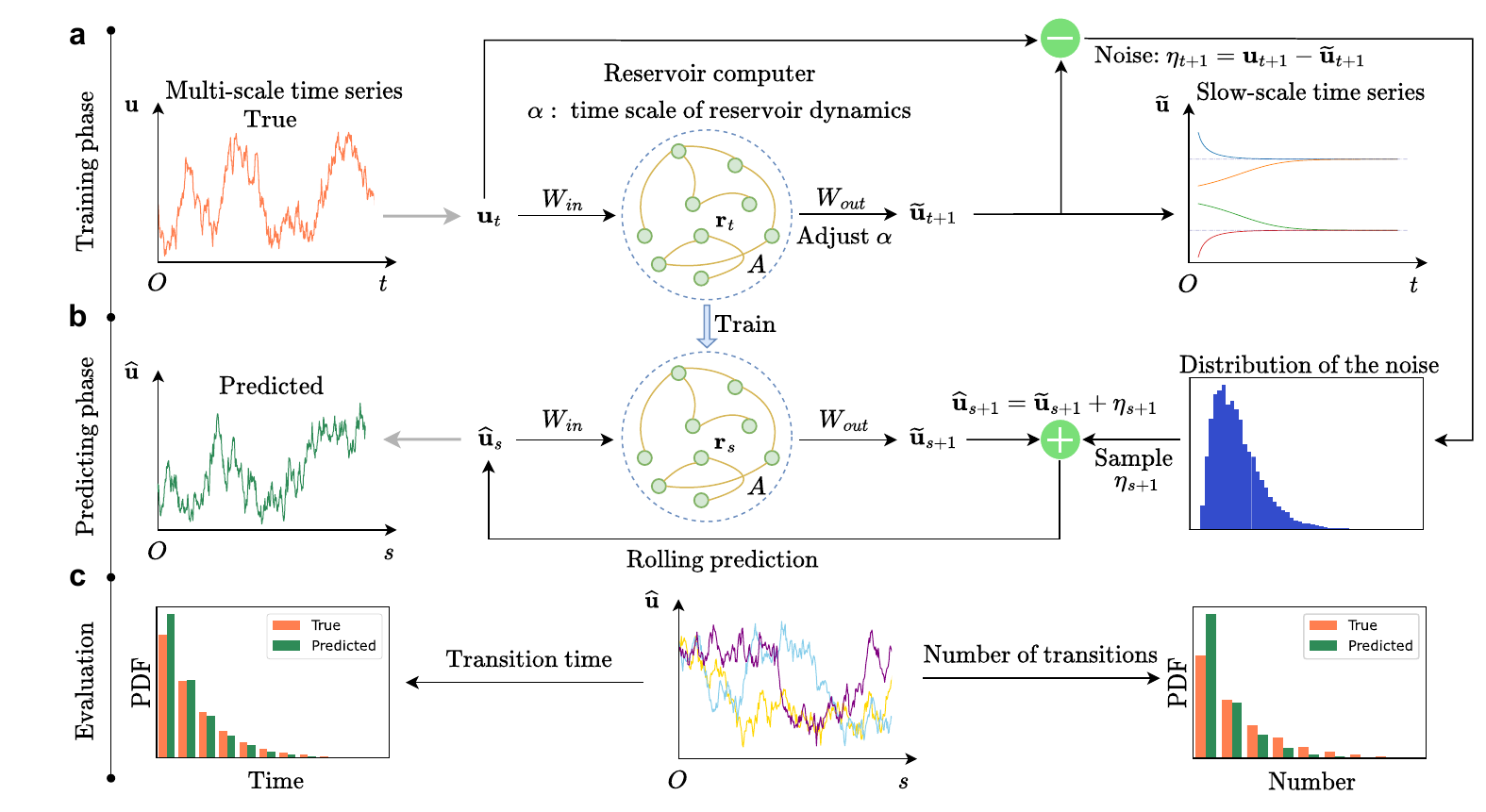}
    
    \caption{
\textbf{Framework of learning noise-induced transitions by multi-scaling reservoir computing.}
(a) The training data is a time series $\mathbf{u}$ with slow and fast time scales, and the fast time-scale part can be considered as noise, causing noise-induced transitions between stable states. In the training phase, at each time step $t$, the reservoir takes into $\mathbf{u}_t$ through a matrix $W_{in}$ and has reservoir state $\mathbf{r}_t$ with a connection matrix $A$. The output matrix $W_{out}$ is trained to fit the output time series to the training data at the next time point. Tuning the hyperparameter $\alpha$ alters the time scale of the output $\widetilde{\mathbf{u}}$, and a properly chosen $\alpha$ leads to a match with the slowly time-scale data. Then, $\mathbf{u}-\widetilde{\mathbf{u}}$ separates the fast time-scale signal $\mathbf{\eta}$ as a noise distribution. (b) In the predicting phase, the $\widehat{\mathbf{u}}_{s}$ is put into the trained reservoir to generate  $\widetilde{\mathbf{u}}_{s+1}$. In the next time step $s+1$, the input $\widehat{\mathbf{u}}_{s+1}$ is the $\widetilde{\mathbf{u}}_{s+1}$ plus the noise $\mathbf{\eta}_{s+1}$ sampled from the separated noise distribution. This process is iterated as a rolling prediction \cite{fang2023reservoir} to generate the time series $\widehat{\mathbf{u}}$. 
(c) The evaluation on the predicted transition statistics. In the middle, different colored lines of $\widehat{\mathbf{u}}$ represent replicates of the predictions. The accuracy is evaluated by the statistics of transition time and the number of transitions. PDF: probability density function.}
    \label{processfigure}
\end{figure}

To demonstrate the effectiveness of the present method, we apply it to two categories of scenarios. One type has the data generated from stochastic differential equations (SDE), and the other has the experimental data \cite{tapia2023enhanced}. For the first category with white noise, it includes a one-dimensional (1D) bistable gradient system, two-dimensional (2D) bistable gradient and non-gradient systems \cite{tang2018escape}, 1D and 2D gradient systems with a tilted potential, a 2D tilted non-gradient system, and a 2D tristable system \cite{belkacemi2021chasing}. The present approach can capture statistics of the transition time and the number of transitions. For the first category with colored noise, we study a 1D bistable gradient system with Lorenz noise \cite{Lim2020Predicting}, and accurately predict the specific transition time, without the assumption of knowing the deterministic part of dynamics as required in \cite{Lim2020Predicting}. 
For the second category, we apply the approach to the protein folding data \cite{tapia2023enhanced}, and explore the least amount of required data for accurate training, which can help reduce the demand for measurements in experiments. 

\section{Results}

\subsection{The framework of multi-scaling reservoir computing}\label{section A}

The reservoir computing has the following scheme \cite{pathak2018model,jaeger2004harnessing}:
 \begin{align}
 \label{RCeq}
\mathbf{r}_{t+1}&=(1-\alpha)\mathbf{r}_t+\alpha\tanh(A\mathbf{r}_{t}+W_{in}\mathbf{u}_t),\\
\mathbf{\widetilde{u}}_{t+1}&=W_{out}\mathbf{r}_{t+1}.
\end{align}
Here, the vector $\mathbf{u}$ is an $n$-dimensional state vector, and the initial condition is $\mathbf{u}_{0}$ with the lower script denoting the time, $W_{in}$ is the input matrix with the values uniformly sampled in $\left[-K_{in}, K_{in} \right]$, $\mathbf{r}$ is the $N$-dimensional reservoir state vector, $A$ is the adjacency matrix of an Erdős-Rényi network with average degree $D$ to describe the reservoir connection between $N$ nodes, and $\rho$ is the spectral radius of $A$. The $\tanh$ represents our activation function for this study. The $\mathbf{\widetilde{u}}$ is the output vector and $W_{out}$ is the output matrix. The $\alpha$ is the leak hyperparameter, representing the time scale \cite{tanaka2021reservoir}. When we reformulate \cref{RCeq} in its continuous form, it provides a more intuitive interpretation of this relationship:
\begin{align}
    \label{RCeq continue}
    \frac{1}{\alpha}\dot{\mathbf{r}}=-\mathbf{r}+\tanh(A\mathbf{r}+W_{in}\mathbf{u}).
\end{align}

In the training phase, only $W_{out}$ is trained to minimize the difference between the output time series and the training data \cite{lukovsevivcius2012practical}. With the regularization term, the loss function is given by
\begin{align}
\label{eqL}
L&=\sum_{t=1}^{T}||\mathbf{u}_{t}-W_{out}\mathbf{r}_{t}||^{2}+\beta||W_{out}||^{2},
\end{align} 
where  $\beta$ is the regression hyperparameter. We then regress the matrix $W_{out}$ by minimizing the loss function (Methods). By stacking the vector of different time points as vector, such as $\mathbf{U}\doteq[\mathbf{u}_{1},\dots,\mathbf{u}_{T}]$ and $\mathbf{R}\doteq[\mathbf{r}_{1},\dots,\mathbf{r}_{T}]$ with $t=1,\dots,T$, it can be rewritten as a compact form:
\begin{align}
\label{Wout}
W_{out}=(\mathbf{U}\mathbf{R}^{\intercal})\cdot (\mathbf{R}\mathbf{R}^{\intercal}+\beta)^{-1},
\end{align} 
where $\intercal$ denotes the transpose. The problem of determining $W_{out}$ is a simple linear regression, which is less computationally expensive than the neural network that requires the back propagation.

The framework for learning noise-induced transitions using multi-scaling reservoir computing is summarized in \autoref{processfigure}. A reservoir acquires a time series $\mathbf{u}$ that contains signals with both fast and slow time scales. Given that $\alpha$ characterizes the time scale of reservoir computer, we search for an appropriate value of $\alpha$ to capture the slow time scale initially. After identifying an appropriate $\alpha$ value, additional searches are conducted to find suitable values for other hyperparameters. This process aims to improve the accuracy of the results and obtain the trained slow-scale model ($W_{out}$). We utilize the trained slow-scale model to separate the noise distribution from the original series. Following this, we use the noise sampled from the separated distribution and employ the trained slow-scale model to do rolling prediction.

In details,  we leverage  the conventional approach \cite{ lukovsevivcius2012practical} to search for appropriate hyperparameters. Given the multiple stable states in the training set, if slow-scale model effectively captures the slowly time-scale dynamics, the trajectories from various initial points (e.g., ten chosen points) should converge to the corresponding stable state. Thus, we tune hyperparameters to reach this property such that the reservoir dynamics matches the slowly time-scale dynamics. Specifically, we first tune the hyperparameter $\alpha$ because it represents the time scale \cite{tanaka2021reservoir}, and then refine the remaining hyperparameters. The effectiveness of these adjustments is assessed by evaluating whether the resulting ten time series converge to their corresponding stable states. In the case of non-convergence, the hyperparameter is adjusted in the opposite direction. When the hyperparameter adjustments do not further improve convergence, we turn to the next hyperparameter \cite{lukovsevivcius2012practical}, as illustrated in \autoref{processfigure}(a).

After finding the proper hyperparameters, we utilize the trained slow-scale model to separate the noise distribution. Within the training phase, at time step $t$, the reservoir accepts the input $\mathbf{u}_{t}$, resulting in an output $\mathbf{\widetilde{u}}_{t+1}$. Then, the noise at time step $t+1$ can be computed as
\begin{align}
 \label{predicting phase1}
\mathbf{\eta}_{t+1}=\mathbf{u}_{t+1}-\mathbf{\widetilde{u}}_{t+1}. 
\end{align}
 Following this, we obtain the noisy time series and the distribution as depicted in \autoref{processfigure}(a). Once the trained slow-scale model and the noise distribution have been obtained, we can implement a rolling prediction (\autoref{processfigure}(b)). In the predicting phase, at time step $s$, the reservoir accepts $\mathbf{\widehat{u}}_{s}$, yielding the output $\mathbf{\widetilde{u}}_{s+1}$. By adding $\mathbf{\eta}_{s+1}$, sampled from the noise distribution, to the output $\mathbf{\widetilde{u}}_{s+1}$ as 
 \begin{align}
 \label{predicting phase2}
\mathbf{\widehat{u}}_{s+1}=\mathbf{\widetilde{u}}_{s+1}+\mathbf{\eta}_{s+1},
\end{align}
 the $\mathbf{\widehat{u}}_{s+1}$ is used as the input for the time step $s+1$. The vector $\mathbf{\widehat{u}}$ is our prediction.

As illustrated in \autoref{processfigure}(c), to validate that the present method accurately captures noise-induced transitions, we compare the prediction to the test data. For white noise, as  Gaussian white noise is memoryless, we quantify the accuracy of our prediction by the statistics of noise-induced transitions, instead of aiming at predicting a single transition, so our focus is on learning the statistics of transition time and the number of transitions from a set of trajectories. For colored noise, we aim to accurately forecast the occurrence of a specific noise-induced transition. 

We next proceed with two categories of examples. One category is data generated from stochastic differential equations, including a 1D bistable gradient system and a 2D bistable non-gradient system with white noise, as well as a 1D bistable gradient system with colored noise. More examples are provided in Supplementary Note: a 1D tilted bistable gradient system (Supplement FIG.~3), a 2D bistable gradient system (Supplement FIG.~4), 2D tilted bistable gradient (Supplement FIG.~5) and non-gradient (Supplement FIG.~6) systems, and a 2D tristable system (Supplement FIG.~7). The second category focuses on experimental data, where we apply the present method to protein folding data \cite{tapia2023enhanced}. We also assess the performance of using a small part of the dataset recorded (\autoref{realdata} and Supplement FIG.~2).

\subsection{Examples}

\begin{table}[H]
  \centering
  \caption{The list of hyperparameters used for the different examples of the main text.}
  \begin{tabular}{@{\extracolsep{4pt}}llllllllll@{\extracolsep{4pt}}}
    \toprule
    \toprule
    Model & $\delta t$ & $T_{\text{train}}$ (Time steps)&$T_{\text{predict}}$ (Time steps) &$N$&$K_{\text{in}}$&$D$& $\rho$& $\alpha$&$\beta$\\
    \midrule
    Example 1 & $0.01$ & $10000$ & $10000 $&$800$ &$4$ &$4$ & $1.2 \times 10^{-3}$&0.2&$1 \times 10^{-8}$ \\
    Example 2 set 1 &$ 0.01$ &$ 8000 $&$ 8000 $&$1000$ &$1.5$ &$3.2$ & $1.3\times 10^{-3}$&0.25&$1 \times 10^{-7}$ \\
    Example 2 set 2 & $0.01$ & $580$ &$300 $& $800$ &$0.996$ &$0.996$ &$0.806$ & $0.065$&$1 \times 10^{-7}$ \\
    Example 3 & $0.002$ & $20000$ & $20000$ &$1200$ &$1$ &$2.2$ & $1.7 \times 10^{-3}$&0.3&$1 \times 10^{-7}$ \\
    Example 4 & real data & $25000$ & $100000$  &$800$ &$0.04$ &$1.8$ & $0.021$&$0.22$ &$1 \times 10^{-6}$\\
    \bottomrule
    \bottomrule
  \end{tabular}
  \label{Hyperparameters}
\end{table}

\subsubsection{A bistable gradient system with white noise} \label{example 1}
\begin{figure}[!htbp]
    \centering
    \includegraphics[width=\textwidth]{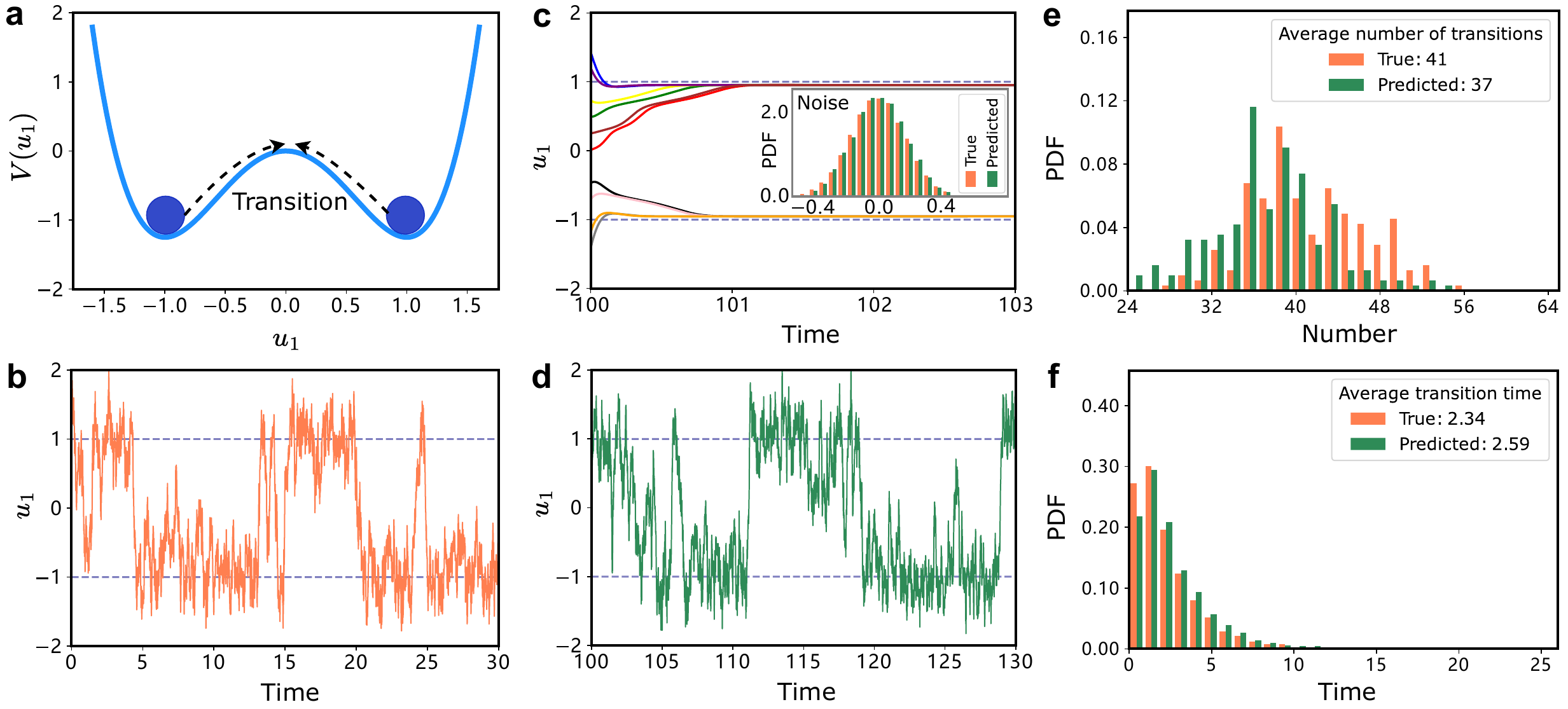}

\caption{\textbf{Capturing stochastic transitions in a bistable gradient system with white noise.} (a) Schematic of noise-induced transitions in the bistable gradient system with Gaussian white noise. (b) Generated time series from \cref{1-d doublewell potential} 
 ($b=5,c=0,\varepsilon=0.3, u_{1}(0)=1.5,\delta t=0.01$) with $t=30$ as ground truth. (c) The trained slow-scale model transforms ten different start points into ten different slowly time-scale series (color lines). Separate the noise distribution in the training phase. (d) Result of  prediction using the slow-scale model and the noise distribution in (c), $t \in [100,130]$. (e) The number of transitions for the test and predicted data matches. Transition refers to the shift from one state to another. The duration of the prediction is $10000\delta t$.
 (f) Histograms of transition time for the test and predicted data. Transition time refers to the interval between the completion of the previous transition and the completion of the current transition.}     
\label{1-D double-well}
\end{figure}

To demonstrate the present method, we first consider data generated from SDE. The continuous-state and continuous-time Markovian stochastic dynamics can be given as
\begin{align}
\dot{\mathbf{u}}=f(\mathbf{u})+\sigma \xi(t),
\label{eq1}
\end{align}
where 
the vector $\dot {\mathbf{u}}$ is the time derivative, the deterministic part of the dynamics is $f(\mathbf{u})$, and $\sigma$ corresponds to the noise strength. The  $\xi(t)$ is a $k$-dimensional Gaussian white noise with $\langle\xi(t)\rangle=0,\langle\xi(t)\xi^\intercal(t^{'})\rangle=\delta(t-t^{'})I_k$, where $I_k$
is the $k$-dimensional identity matrix, $\delta(t-t^{'})$ is the Dirac $\delta$ function, and $\langle\cdots\rangle$ represents the noise average. 

As a first example, we consider a 1D bistable gradient system with white noise \cite{forgoston2018primer}, 
\begin{align}
\label{1-d doublewell potential}
    \dot{u}_1=-b(-u_1+u_1^3+c)+\sqrt{2\varepsilon b}\xi_{1}(t),\enspace t \ge 0,
\end{align}
with a Gaussian white noise $\xi_{1}(t)$. The parameter $b$ denotes the strength of the diffusion coefficient, $\varepsilon$ is the noise strength, and $c$ controls the tilt of the two potential wells. The system has noise-induced transitions between the two potential wells as illustrated in \autoref{1-D double-well}(a). We generated a time series with a duration of $20000\delta t$, with the training set $t\in[0,100]$ and the predicting set $t\in[100,200]$. \autoref{1-D double-well}(b)  shows the first $3,000\delta t$ of the training set.
  
In the training phase, the tuning of hyperparameters for slow-scale model is performed as outlined in \autoref{section A}, utilizing ten initial points to search for proper hyperparameters. \autoref{1-D double-well}(c) illustrates the process, after finding the proper hyperparameters listed in \autoref{Hyperparameters} (Example 1), we obtain the trained slow-scale model. We use the generated slowly time-scale series to separate the noise distribution, as demonstrated in \autoref{processfigure}(a). We employ the trained slow-scale model and the separated noise distribution for rolling prediction. The first $3000\delta t$ of the prediction is illustrated in \autoref{1-D double-well}(d). The prediction series demonstrates similar dynamics to the test data, including noise-induced transitions. 

In the evaluation, we generated  $100$ replicates of time series from \cref{1-d doublewell potential}, trained the model, and produced $100$ time series separately. Then, we compare the statistics of the noise-induced transitions for these two sets of time series, e.g., the number of transitions over $10000\delta t$ (\autoref{1-D double-well}(e)) and the transition time (\autoref{1-D double-well}(f)). The match between the test and predicted data demonstrates the effectiveness of our approach in capturing noise-induced transition dynamics.


\begin{figure}[htbp]
    \centering
    \includegraphics[width=\textwidth]{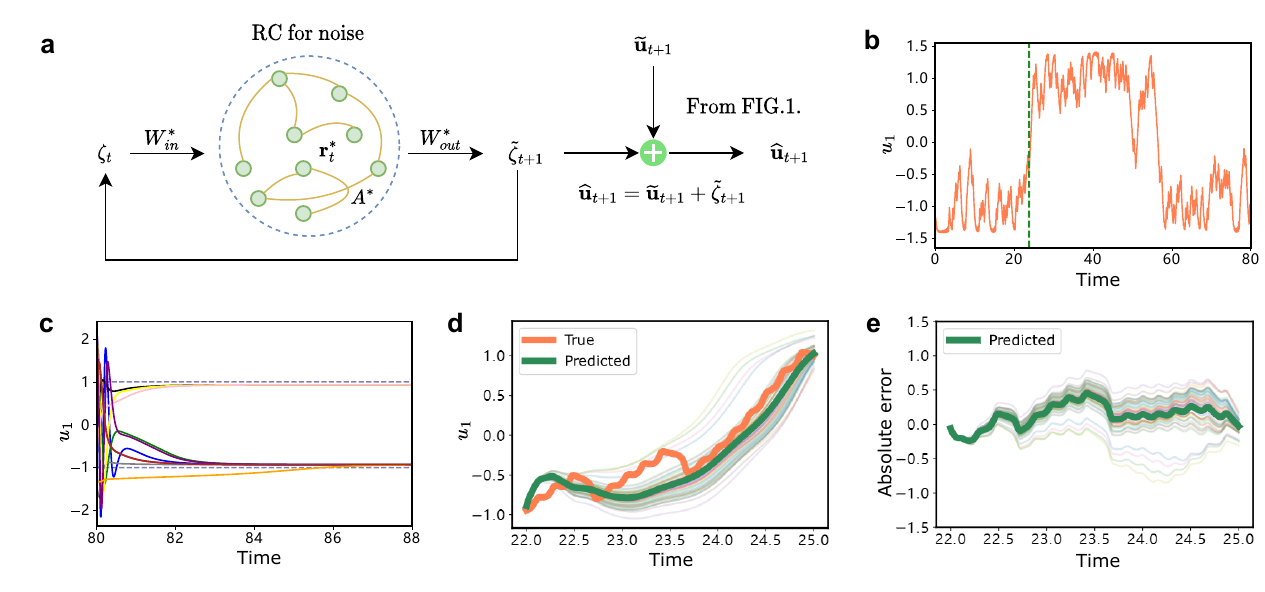}
    \caption{\textbf{Predicting the accurate transition time series for a bistable gradient system with colored noise.} The system is the same as \cref{1-D doublewell with lorenz} \cite{Lim2020Predicting}.  (a) The flowchart of predicting stochastic transitions with colored noise. The process for obtaining noise $\zeta_{t}$ follows that in \autoref{processfigure}(a), and a second reservoir takes into $\zeta_{t}$ through matrix $W_{in}^{*}$ and has reservoir states $\mathbf{r}^{*}_{t}$  with a connection matrix $A^{*}$. The output matrix $W_{out}^{*}$  is trained to learn noise. Using the trained slow-scale model to predict noise-induced transitions. (b) Generated time series ($x(0)=y(0)=z(0)=1, b=1,  c=0, \psi=0.08 ,\varepsilon=0.5, u_{1}(0)=-1.5, \delta t=0.01$) with $8000\delta t$, where a noise-induced transition occurs in $t\in[22,25]$ marked by the green dashed line. The noise data from $580\delta t$ before the stochastic transition at $t=22$ is applied to predict the noisy time series in $t\in [22,25]$. (c) The trained slow-scale model transforms ten different start points into ten different slowly time-scale series (color lines). Separate the noisy time series in the training phase. (d) By repeating the process in (a) $50$ times with the same hyperparameters, $50$ predicted $u_1(t)$ are obtained (fainter lines). The averaged predicted time series (thick green) matches the test data (coral). (e) Absolute error of the predicted $50$ time series and its mean value (thick green).}
    \label{fig:1Ddouble-well Lorenz}
\end{figure}

\subsubsection{A bistable gradient system with colored noise}
\label{example 2}

The prediction on a single stochastic transition may become possible when the system has colored noise. To demonstrate that the present method is applicable to such cases, we consider a system \cref{1-D doublewell with lorenz,Lorenz in x,Lorenz in y,Lorenz in z} studied in \cite{Lim2020Predicting}, where their method relies on the assumption of knowing the deterministic part of the equation in prior. In contrast to their method, we do not assume any prior knowledge of the deterministic part of the dynamical system and directly learn the deterministic part and noise (\autoref{fig:1Ddouble-well Lorenz}(a)), enabling a prediction in a model-free manner.

The system is a 1D bistable gradient system, as illustrated in \autoref{fig:1Ddouble-well Lorenz}(b):  
\begin{align}
    \label{1-D doublewell with lorenz}
    \dot{u}_1&=-b(-u_1+u_1^3+c)+\frac{\psi}{\varepsilon}y,\\
    \label{Lorenz in x}
    \dot{x}&=\frac{10}{\varepsilon^2}(y-x),\\
    \label{Lorenz in y}
    \dot{y}&=\frac{1}{\varepsilon^2}(28 x-xz-y),\\
    \label{Lorenz in z}
    \dot{z}&=\frac{1}{\varepsilon^2}(xy-\frac{8}{3} z).
\end{align} The parameter $b$ denotes the strength
of the diffusion coefficient, $\varepsilon$ corresponds to the noise strength, $\psi$ controls the influence of the noise on the slow-scale dynamics, and $c$ controls the tilt of the two potential wells. The noise $(x, y, z)$  is modeled by a dynamical system, the Lorenz-63 model \cite{lorenz1963deterministic}. The system has stochastic transitions between the two potential wells under the Lorenz noise.

To demonstrate the present method, we consider the time series for $8000\delta t$ (\autoref{fig:1Ddouble-well Lorenz}(b)), where a stochastic transition is indicated prior to the green dashed line. In the training phase, we  obtain slow-scale model to learn the deterministic part (\autoref{fig:1Ddouble-well Lorenz}(c)), and to separate noise. Hyperparameters are listed in \autoref{Hyperparameters} (Example 2 set 1). In the predicting phase, accurately forecasting the stochastic transitions requires to predict the noisy time series. Thus, we utilize a second reservoir (\autoref{fig:1Ddouble-well Lorenz}(a)) to learn the previously separated noise during the training phase. The hyperparameters for the noisy time series are listed in \autoref{Hyperparameters} (Example 2 set 2). With the deterministic component slow-scale model, we execute a rolling prediction to predict transition. 

To evaluate the accuracy of prediction, as in \cite{Lim2020Predicting}, we applied the same hyperparameters to conduct $50$ predictions. These predictions were then used alongside trained slow-scale model for $50$ times rolling prediction.  The average of the $50$ predictions outcomes closely approximates the actual time series (\autoref{fig:1Ddouble-well Lorenz}(d)). Furthermore, \autoref{fig:1Ddouble-well Lorenz}(e) shows a near-zero average absolute error between the $50$ predictions and the actual time series, indicating a high accuracy. These results demonstrate that the present approach requires no assumptions on knowing the deterministic part, underscoring its potential in predicting a single stochastic transition under colored noise.


\subsubsection{A bistable non-gradient system} \label{example 3} 
\begin{figure}[htbp]
    \centering
    \includegraphics[width=\textwidth]{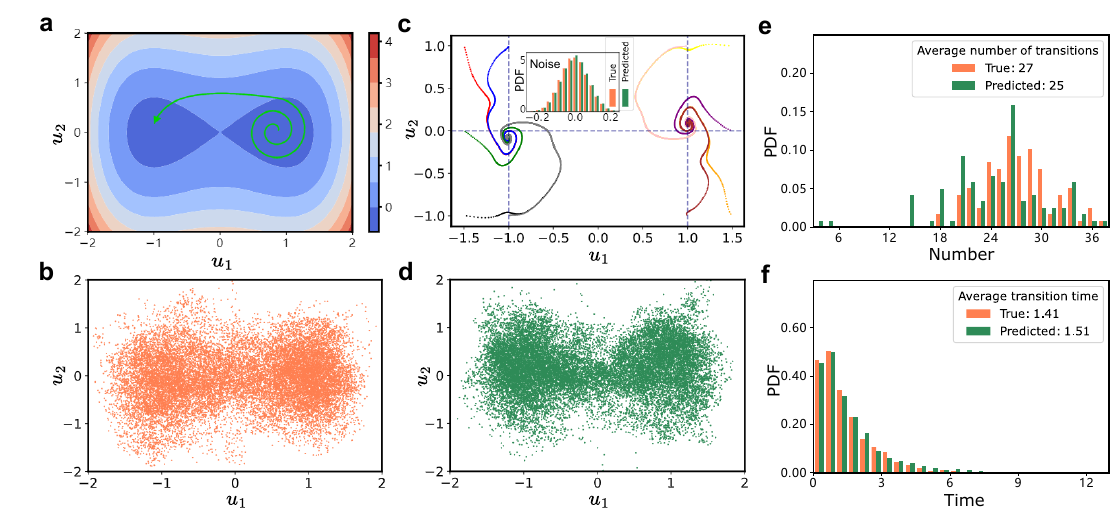}
    \caption{\textbf{Learning noise-induced transitions in a bistable non-gradient system.} (a) 
Schematic of noise-induced transitions in the 2D bistable non-gradient system with Gaussian white noise.
(b) Generated time series from \cref{2-D doublewell in x,2-D doublewell in y} 
 ($a=b=5, c=0, \varepsilon_{1}=\varepsilon_{2}=0.3, u_{1}(0)=0,u_{2}(0)=2,\delta t=0.002$) with $t=40$ as ground truth.  (c) The trained slow-scale model transforms ten different start points into ten different slowly time-scale series (color lines), $t \in [40,80]$. Separate the noise distribution. (d) Result of  prediction using the slow-scale model and the noise distribution in (c). (e) The number of transitions for the 100 replicates simulated in $t \in[40,80]$ and the 100 generated matches. The transition refers to the time series in the $u_1$-direction that crosses the zero point and remains either non-negative or non-positive for $50\delta t$.
(f) Histograms of transition time for the test and predicted data. When a transition occurs within the system, the transition time is defined as the interval between two consecutive zero crossings in the $u_{1}$ direction. }
    \label{2d doublewell}
\end{figure}
We next focus on investigating whether the noise-induced transitions can be predicted by the present method for the 2D non-gradient systems. We consider a bistable non-gradient system \cite{tang2018escape}:
\begin{align}
\label{2-D doublewell in x}
    \dot{u}_1&=-b(-u_1+u_1^3+c)-au_2+\sqrt{2\varepsilon_1 b}\xi_1(t),\enspace t \ge 0,\\
    \label{2-D doublewell in y}
     \dot{u}_2&=a(-u_1+u_1^3+c)-bu_2+\sqrt{2\varepsilon_2 b}\xi_2(t),\enspace t \ge 0,
\end{align}
with Gaussian white noise $\xi_{1}(t)$ and $\xi_{2}(t)$. In this system, $b$ is the diffusion coefficient, $a$ represents the strength of the non-detailed balance part, $\varepsilon_1$ and $ \varepsilon_2$ are the noise strength, and $c$ controls the tilt of the potential. The system has noise-induced transitions between 
the two potential wells under the noise as illustrated in \autoref{2d doublewell}(a). The presence of a non-detailed balance introduces a rotational component to the time series, which adds difficulty to the prediction.

In the training phase, we generated a series consisting of $40000\delta t$. The training set is $t\in[0,40]$, and the predicting set is $t\in[40,80]$. \autoref{2d doublewell}(b) displays the training set. As the method in \autoref{section A}, the deterministic part is plotted in \autoref{2d doublewell}(c). A proper set of hyperparameters is listed in \autoref{Hyperparameters} (Example 3). We observe that the generated time series starting from the ten initial points converge to two potential wells, where the time series has rotational dynamics. In the predicting phase, we perform rolling prediction within $t\in [40,80]$ (\autoref{2d doublewell}(d)).

In the evaluation, we predict $100$ replicates of the time series, and compare them to $100$ replicates simulated from \cref{2-D doublewell in x,2-D doublewell in y}.  \autoref{2d doublewell}(e) presents histograms of the number of transitions for the $100$ predicted time series and $100$ test time series, in $t\in[40,80]$.  \autoref{2d doublewell}(f) presents histograms of transition time for the $100$ predicted time series and the $100$ test series, in $t\in[40,80]$. The results demonstrate that, for the 2D bistable non-gradient system, the present method accurately learns the dynamics and yields precise estimations on the number of transitions and transition time.

\begin{figure}[htbp]
    \centering    \includegraphics[width=\textwidth]{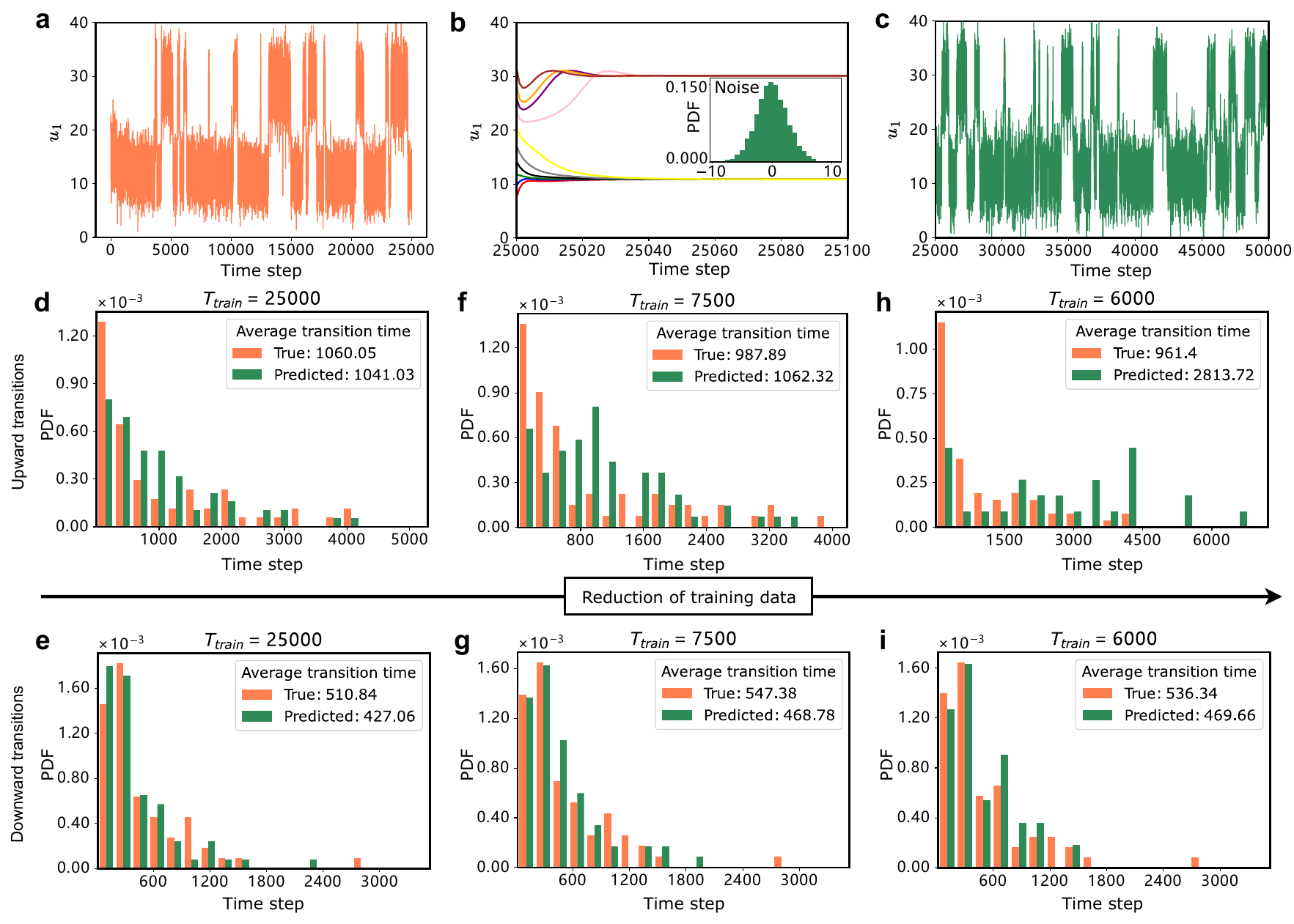}
    \caption{\textbf{Learning the stochastic transitions from the experimental data of protein folding.} The $u_1$ represents the end-to-end length of the protein.  We refer to transitions from around $u_1=15$ to around $u_1=30$ as upward transitions, and vice versa as downward transitions. The right-pointing arrow: reduction of training data.
    (a) Time series of the training set ($0-25000$ time steps). (b) The trained slow-scale model transforms ten different start points into ten different slowly time-scale series (color lines). Separate noise distribution in the training phase.  (c) Result of prediction using the trained slow-scale model and separated noise in (b) during time steps $25000-50000$.  (d) Histograms of upward transition time for the prediction and the true data, where the length of the training set $(T_{train})$ is $25000$ time steps. Transition time refers to the interval between the completion of the previous transition and the completion of the current transition. (e) Histograms of downward transition time for the prediction and the true data, with $T_{train}=25000$ time steps. (f-i) Similar histograms of upward and downward transition time as in (d) and (e), and with different lengths of the training sets, $T_{train}=7500$ time steps for (f) and (g), and $T_{train}=6000$ time steps for (h) and (i). This match demonstrates that the present method can still make predictions even when reducing the training length to $T_{train}=7500$.}
    \label{realdata}
\end{figure}

\subsubsection{Experimental data of protein folding}
\label{example 4}

To demonstrate that the present method can learn the functional role of noise in stochastic transitions of experimental data, we apply the present method to the protein folding data \cite{tapia2023enhanced}. The protein can escape the free energy barrier from one stable state to another stable state, variations in end-to-end length correspond to different free energy, which can result in
the protein being either in an unfolded or native conformation. This process can be regarded as a noise-induced transition. \autoref{realdata}(a) shows the training data, where there are transitions between two stable states.

In the training phase, with $T_{train}=25000$ time steps,  we obtain the trained slow-scale model and ten different slowly time-scale series with the separated noise distribution (\autoref{realdata}(b)). The proper hyperparameters are listed in \autoref{Hyperparameters} (Example 4). In the predicting phase, we employ the trained slow-scale model and the separated noise distribution to do rolling prediction for $100000$ time steps. The first $25000$ time steps of prediction are plotted in \autoref{realdata}(c), showing the transitions between stable states and the asymmetric dynamics.

In experiments, the available data is often limited, and it is essential to determine the minimum amount of data required. Thus, we reduce the amount of training data to $7500$ and $6000$ time steps separately. We generate prediction for $100000$ time steps and then compare this prediction to the true data to evaluate the impact of data length on prediction accuracy. The results in \autoref{realdata}(d-e) and \autoref{realdata}(f-g)  demonstrate 
that the present method can learn the dynamics of protein folding from the data with around $7500$ time steps. When $T_{train}=6000$ time steps, \autoref{realdata}(h-i) show a larger error between the predicted and true transition time. This suggests that $7500$ time steps approximate the minimum data requirement for the present method in this system, 
such that the behavior of protein folding can be effectively learnt and simulated for more time steps. Therefore, the present approach holds promising potential for streamlining the workload of experimentalist by allowing to  learn protein folding dynamics from a small dataset. 

\section{Discussion}
 
We have provided a general framework for noise-induced transitions that is solely based on data. The present method exhibits considerable improvements compared with the previous work \cite{Lim2020Predicting}: A crucial distinction is that we can obtain the deterministic part of dynamics without any prior knowledge about underlying deterministic equation. We have applied the method to examples from stochastic differential equations and the  experimental data, where the method can accurately  learn noise-induced transitions and estimate transition statistics in various systems.

The choice of hyperparameters affects the training. For example, the hyperparameter $\alpha$ plays an important role: the larger $\alpha$ corresponds to the time series with fast time scale, while the smaller $\alpha$ leads to the slow time scale  \cite{tanaka2021reservoir}. We utilize this characteristic to search for $\alpha$ to match with the slow dynamics and separate noise.  For the example with colored noise (\autoref{example 2}), the $\alpha$ for noise is smaller than that for the deterministic part (\autoref{Hyperparameters}). 
We find that using a smaller $\alpha$ to learn the noisy time series can better predict transitions than using a larger $\alpha$. As the fast and chaotic noisy time series is often difficult to predict, using a small $\alpha$ leads to smoother noisy time series. While this selection may sacrifice some details of noise, it enhances the ability to capture the major trend of noise and thus improves transition prediction. In asymmetric systems, such as tilted bistable gradient systems, the two distinct potential wells of the system exhibit different time scales. Consequently, we need to employ two different sets of hyperparameters (Supplement FIGs. 3, 5, and 6). 
If a time series is generated from an asymmetric system, our multi-scaling reservoir computing approach can identify the tilted dynamics and accurately simulate two types of transitions, even without any prior information about its tilt and noise strength.

The effectiveness of learning slowly time-scale series can also be influenced by other hyperparameters \cite{lukovsevivcius2012practical}. In our protocol, we utilize ten initial points to search for appropriate hyperparameters. Given the inherent complexity of neural network and random matrices in reservoir computer, developing a universal and systematic strategy for selecting predefined hyperparameters is a significant challenge \cite{PhysRevResearch.1.033056,gauthier2021next}. 
As potential ways of improvements, the application of Bayesian optimization \cite{yperman2016bayesian} and simulated annealing \cite{ren2022global} can be used to facilitate the search for hyperparameters. Additionally, the convergence speed of the ten slowly time-scale series may have some discrepancies compared with the actual dynamics.  As a result, the noise distribution separated by the slow-scale model during the training phase may exhibit either lower or higher intensity compared with the actual noise distribution. In this case, we can employ a noise factor to magnify or reduce separated noise, aiming to predict transitions more accurately.  To verify this idea, we amplify the sampled noise by a factor of $1.1$ based on \autoref{example 1}, which improves the accuracy of the predictions (Supplement FIG.~1). 

For experimental data, we have shown the feasibility of predicting the experimental time series from a small dataset, as exemplified in the analysis of the protein folding data \cite{tapia2023enhanced}. We can capture the transitions and the tilted potential dynamics. In this example, over short timescales, the protein samples a local equilibrium involving the native and unfolded conformations. However, if the measurement time is significantly extended, previously inaccessible regions separated by high-energy barriers can be explored. Consequently, in order to capture a wider variety of protein folding transitions, it may be necessary to use longer training set, which would result in higher computational costs. 
Additionally, we observed tilted dynamics in the time series of protein folding. Even so, we can learn both the upward and downward transitions by using only one set of hyperparameters. This suggests that in this system the upward and downward transition time scales might not differ significantly. The noise strength drives the system to cross one energy barrier more frequently while encountering difficulty in crossing another. Therefore, when dealing with a tilted dynamics time series, 
we can utilize two distinct sets of hyperparameters to learn the different time scales.

The recent Python library for machine learning dynamical models from time series, Deeptime \cite{hoffmann2021deeptime}, has not directly handled the stochastic transitions. There are three categories of methods in Deeptime related to our work. The first category is the deep dimension reduction, such as deep Koopman networks. These methods can find stable states from time series and identify proper time scales to separate slow and fast
dynamics \cite{li2023learning}. However, they have not achieved to predict stochastic transitions between stable states. The second category method is the sparse identification of nonlinear dynamics (SINDy). Although it can identify nonlinear dynamics from noisy data \cite{kaheman2022automatic}, it mainly regards noise as a disturbance and is not designed for learning noise-induced phenomena. The third category comprises Markov state models and hidden Markov models. These methods are applicable to stochastic processes with discrete states, whereas the present study focuses on stochastic processes with continuous states.

There are more potential applications of the present method in various domains. In physics, such as in Nagumo's tunnel diode neuron model, the inclusion of a nonlinear resistor \cite{semenov2016noise} can lead to noise-induced transitions. We can apply the present method to learn these transitions between different voltage and current stable states. The present approach may also be extended 
to analyze transition of trajectories between different dynamical phases of spins \cite{casert2022learning,tang2022solving}, and to study the switches of different meander states in climate change \cite{vanden2013data}. Furthermore, we can further extend the approach for examples involving a wider variety of noise types, such as band-limited thermal noise, power law noise, shot noise, and impulsive noise, where the conditional generative adversarial network  \cite{wunderlich2022data} may be employed to model these noise.

\clearpage
\section{Methods}

In this section, we reformulate the loss function to derive the expression for the output matrix $W_{out}$ and discuss the hyperparameters in the present method. The loss function is given as \cref{eqL}. In detail, we should write the loss function as a sum from all the parameters to do linear regression. Then,  the regression becomes simply a sum of vectors:
\begin{align}
L&=\sum_{t=1}^{T}[||\mathbf{u}_{t}-W_{out}\mathbf{r}_{t}||^{2}+\beta||W_{out}||^{2}]
\notag\\&=\sum_{t=1}^{T}[(\mathbf{u}_{t}-W_{out}\mathbf{r}_{t})^{\intercal}(\mathbf{u}_{t}-W_{out}\mathbf{r}_{t})+\beta||W_{out}||^{2}]
\notag\\&=\sum_{t=1}^{T}[(\mathbf{u}_{t})^{\intercal}\mathbf{u}_{t}-(W_{out}\mathbf{r}_{t})^{\intercal}\mathbf{u}_{t}-(\mathbf{u}_{t})^{\intercal}W_{out}\mathbf{r}_{t}+(W_{out}\mathbf{r}_{t})^{\intercal}W_{out}\mathbf{r}_{t}+\beta||W_{out}||^{2}].
\end{align}
 As the loss function is convex (prove that the zero gradient is indeed the local minimum, one needs to differentiate once more to obtain the Hessian matrix and show that it is positive definite. This is provided by the Gauss-Markov theorem), the optimum solution lies at the zero gradient by
\begin{align}
\partial_{W_{out}}L&=\sum_{t=1}^{T}[-2(\mathbf{r}_{t})^{\intercal}\mathbf{u}_{t}+2(\mathbf{r}_{t})^{\intercal}W_{out}\mathbf{r}_{t}+2\beta W_{out}]=0,
\end{align}
which leads to the regression:
\begin{align}
W_{out}=\sum_{t=1}^{T}\left[(\mathbf{u}_{t})\cdot(\mathbf{r}_{t})^{\intercal}\right]\cdot [(\mathbf{r}_{t})\cdot(\mathbf{r}_{t})^{\intercal}+\beta]^{-1},
\end{align}
where we have neglected the notation of identity matrix and identify vector.
By stacking the vector of different time points as vector, it can be rewritten as a compact form \cref{Wout}.

There are six hyperparameters in the present method. The variable $N$ represents the number of reservoir nodes, which determines reservoir size. In most instances, performance improves with larger reservoir \cite{verzelli2022learning}. However, using large reservoir might lead to overfitting, requiring the application of suitable regularization techniques \cite{lukovsevivcius2012practical}. The hyperparameter $K_{in}$ represents the scaling factor for the input matrix $W_{in}$. The average degree of the reservoir connection network is denoted by $D$, and we choose the connection matrix $A$ to be sparse \cite{jaeger2001echo}. This approach stems from the intuition that decoupling the state variables can result in a richer encoding of the input signal \cite{verzelli2022learning}. The spectral radius of the reservoir connection network, denoted as $\rho$, represents a critical characteristic of the dynamics of the reservoir state. Notably, it affects both the nonlinearity of the reservoir and its capacity to encode past inputs in its state \cite{verzelli2022learning,du2023inferring}. The $\alpha$ represents the leak parameter, it represents the time scale \cite{tanaka2021reservoir}. The $\beta$ is the regularization parameter \cite{lukovsevivcius2012practical}.

\textbf{Data availability:}
The authors declare that the data supporting this study are available within the paper.

\textbf{Code availability:}
A pytorch implementation of the present algorithm will be publicly available upon the acceptance of the manuscript.

\section*{Acknowledgments}
We acknowledge Xingang Wang for helpful discussions. 
We thank Rafael Tapia-Rojo for sharing the data of protein folding. 
This work is supported by Project 12322501, 12105014 of National Natural Science Foundation of China. 
Y.T. acknowledges the start-up research funding (28705-310432101) by Beijing Normal University. 
The HPC is supported by Interdisciplinary Intelligence SuperComputer Center of Beijing Normal University, Zhuhai.

\section*{Author contributions}
Y.T. had the original idea for this work, Z.Q.L. and Z.F.L. performed the study, and all authors contributed to the preparation of the manuscript.

\bibliography{bib}

\begin{thebibliography}{52}%
\makeatletter
\providecommand \@ifxundefined [1]{%
 \@ifx{#1\undefined}
}%
\providecommand \@ifnum [1]{%
 \ifnum #1\expandafter \@firstoftwo
 \else \expandafter \@secondoftwo
 \fi
}%
\providecommand \@ifx [1]{%
 \ifx #1\expandafter \@firstoftwo
 \else \expandafter \@secondoftwo
 \fi
}%
\providecommand \natexlab [1]{#1}%
\providecommand \enquote  [1]{``#1''}%
\providecommand \bibnamefont  [1]{#1}%
\providecommand \bibfnamefont [1]{#1}%
\providecommand \citenamefont [1]{#1}%
\providecommand \href@noop [0]{\@secondoftwo}%
\providecommand \href [0]{\begingroup \@sanitize@url \@href}%
\providecommand \@href[1]{\@@startlink{#1}\@@href}%
\providecommand \@@href[1]{\endgroup#1\@@endlink}%
\providecommand \@sanitize@url [0]{\catcode `\\12\catcode `\$12\catcode
  `\&12\catcode `\#12\catcode `\^12\catcode `\_12\catcode `\%12\relax}%
\providecommand \@@startlink[1]{}%
\providecommand \@@endlink[0]{}%
\providecommand \url  [0]{\begingroup\@sanitize@url \@url }%
\providecommand \@url [1]{\endgroup\@href {#1}{\urlprefix }}%
\providecommand \urlprefix  [0]{URL }%
\providecommand \Eprint [0]{\href }%
\providecommand \doibase [0]{https://doi.org/}%
\providecommand \selectlanguage [0]{\@gobble}%
\providecommand \bibinfo  [0]{\@secondoftwo}%
\providecommand \bibfield  [0]{\@secondoftwo}%
\providecommand \translation [1]{[#1]}%
\providecommand \BibitemOpen [0]{}%
\providecommand \bibitemStop [0]{}%
\providecommand \bibitemNoStop [0]{.\EOS\space}%
\providecommand \EOS [0]{\spacefactor3000\relax}%
\providecommand \BibitemShut  [1]{\csname bibitem#1\endcsname}%
\let\auto@bib@innerbib\@empty
\bibitem [{\citenamefont {Horsthemke}\ and\ \citenamefont
  {Lefever}(2006)}]{horsthemke1984noise}%
  \BibitemOpen
  \bibfield  {author} {\bibinfo {author} {\bibfnamefont {W.}~\bibnamefont
  {Horsthemke}}\ and\ \bibinfo {author} {\bibfnamefont {R.}~\bibnamefont
  {Lefever}},\ }\href@noop {} {\emph {\bibinfo {title} {Noise-Induced
  Transitions: Theory and Applications in Physics, Chemistry, and Biology}}},\
  \bibinfo {edition} {2nd}\ ed.\ (\bibinfo  {publisher} {Springer-Verlag,
  Berlin},\ \bibinfo {year} {2006})\BibitemShut {NoStop}%
\bibitem [{\citenamefont {Semenov}\ \emph {et~al.}(2016)\citenamefont
  {Semenov}, \citenamefont {Neiman}, \citenamefont {Vadivasova},\ and\
  \citenamefont {Anishchenko}}]{semenov2016noise}%
  \BibitemOpen
  \bibfield  {author} {\bibinfo {author} {\bibfnamefont {V.~V.}\ \bibnamefont
  {Semenov}}, \bibinfo {author} {\bibfnamefont {A.~B.}\ \bibnamefont {Neiman}},
  \bibinfo {author} {\bibfnamefont {T.~E.}\ \bibnamefont {Vadivasova}},\ and\
  \bibinfo {author} {\bibfnamefont {V.~S.}\ \bibnamefont {Anishchenko}},\
  }\bibfield  {title} {\bibinfo {title} {Noise-induced transitions in a
  double-well oscillator with nonlinear dissipation},\ }\href
  {https://doi.org/https://doi.org/10.1103/PhysRevE.93.052210} {\bibfield
  {journal} {\bibinfo  {journal} {Phys. Rev. E}\ }\textbf {\bibinfo {volume}
  {93}},\ \bibinfo {pages} {052210} (\bibinfo {year} {2016})}\BibitemShut
  {NoStop}%
\bibitem [{\citenamefont {Assaf}\ \emph {et~al.}(2011)\citenamefont {Assaf},
  \citenamefont {Roberts},\ and\ \citenamefont
  {Luthey-Schulten}}]{PhysRevLett.106.248102}%
  \BibitemOpen
  \bibfield  {author} {\bibinfo {author} {\bibfnamefont {M.}~\bibnamefont
  {Assaf}}, \bibinfo {author} {\bibfnamefont {E.}~\bibnamefont {Roberts}},\
  and\ \bibinfo {author} {\bibfnamefont {Z.}~\bibnamefont {Luthey-Schulten}},\
  }\bibfield  {title} {\bibinfo {title} {Determining the stability of genetic
  switches: Explicitly accounting for mrna noise},\ }\href
  {https://doi.org/10.1103/PhysRevLett.106.248102} {\bibfield  {journal}
  {\bibinfo  {journal} {Phys. Rev. Lett.}\ }\textbf {\bibinfo {volume} {106}},\
  \bibinfo {pages} {248102} (\bibinfo {year} {2011})}\BibitemShut {NoStop}%
\bibitem [{\citenamefont {Jafarpour}\ \emph {et~al.}(2015)\citenamefont
  {Jafarpour}, \citenamefont {Biancalani},\ and\ \citenamefont
  {Goldenfeld}}]{PhysRevLett.115.158101}%
  \BibitemOpen
  \bibfield  {author} {\bibinfo {author} {\bibfnamefont {F.}~\bibnamefont
  {Jafarpour}}, \bibinfo {author} {\bibfnamefont {T.}~\bibnamefont
  {Biancalani}},\ and\ \bibinfo {author} {\bibfnamefont {N.}~\bibnamefont
  {Goldenfeld}},\ }\bibfield  {title} {\bibinfo {title} {Noise-induced
  mechanism for biological homochirality of early life self-replicators},\
  }\href {https://doi.org/10.1103/PhysRevLett.115.158101} {\bibfield  {journal}
  {\bibinfo  {journal} {Phys. Rev. Lett.}\ }\textbf {\bibinfo {volume} {115}},\
  \bibinfo {pages} {158101} (\bibinfo {year} {2015})}\BibitemShut {NoStop}%
\bibitem [{\citenamefont {Qian}(2002)}]{qian2002discrete}%
  \BibitemOpen
  \bibfield  {author} {\bibinfo {author} {\bibfnamefont {H.}~\bibnamefont
  {Qian}},\ }\bibfield  {title} {\bibinfo {title} {From discrete protein
  kinetics to continuous brownian dynamics: A new perspective},\ }\href
  {https://doi.org/https://doi.org/10.1110/ps.18902} {\bibfield  {journal}
  {\bibinfo  {journal} {Protein Sci.}\ }\textbf {\bibinfo {volume} {11}},\
  \bibinfo {pages} {1} (\bibinfo {year} {2002})}\BibitemShut {NoStop}%
\bibitem [{\citenamefont {Tapia-Rojo}\ \emph {et~al.}(2023)\citenamefont
  {Tapia-Rojo}, \citenamefont {Mora}, \citenamefont {Board}, \citenamefont
  {Walker}, \citenamefont {Boujemaa-Paterski}, \citenamefont {Medalia},\ and\
  \citenamefont {Garcia-Manyes}}]{tapia2023enhanced}%
  \BibitemOpen
  \bibfield  {author} {\bibinfo {author} {\bibfnamefont {R.}~\bibnamefont
  {Tapia-Rojo}}, \bibinfo {author} {\bibfnamefont {M.}~\bibnamefont {Mora}},
  \bibinfo {author} {\bibfnamefont {S.}~\bibnamefont {Board}}, \bibinfo
  {author} {\bibfnamefont {J.}~\bibnamefont {Walker}}, \bibinfo {author}
  {\bibfnamefont {R.}~\bibnamefont {Boujemaa-Paterski}}, \bibinfo {author}
  {\bibfnamefont {O.}~\bibnamefont {Medalia}},\ and\ \bibinfo {author}
  {\bibfnamefont {S.}~\bibnamefont {Garcia-Manyes}},\ }\bibfield  {title}
  {\bibinfo {title} {Enhanced statistical sampling reveals microscopic
  complexity in the talin mechanosensor folding energy landscape},\ }\href
  {https://doi.org/https://doi.org/10.1038/s41567-022-01808-4} {\bibfield
  {journal} {\bibinfo  {journal} {Nat. Phys.}\ }\textbf {\bibinfo {volume}
  {19}},\ \bibinfo {pages} {52} (\bibinfo {year} {2023})}\BibitemShut {NoStop}%
\bibitem [{\citenamefont {H\"anggi}\ \emph {et~al.}(1990)\citenamefont
  {H\"anggi}, \citenamefont {Talkner},\ and\ \citenamefont
  {Borkovec}}]{RevModPhys.62.251}%
  \BibitemOpen
  \bibfield  {author} {\bibinfo {author} {\bibfnamefont {P.}~\bibnamefont
  {H\"anggi}}, \bibinfo {author} {\bibfnamefont {P.}~\bibnamefont {Talkner}},\
  and\ \bibinfo {author} {\bibfnamefont {M.}~\bibnamefont {Borkovec}},\
  }\bibfield  {title} {\bibinfo {title} {Reaction-rate theory: fifty years
  after kramers},\ }\href {https://doi.org/10.1103/RevModPhys.62.251}
  {\bibfield  {journal} {\bibinfo  {journal} {Rev. Mod. Phys.}\ }\textbf
  {\bibinfo {volume} {62}},\ \bibinfo {pages} {251} (\bibinfo {year}
  {1990})}\BibitemShut {NoStop}%
\bibitem [{\citenamefont {Tang}\ \emph {et~al.}(2023)\citenamefont {Tang},
  \citenamefont {Weng},\ and\ \citenamefont {Zhang}}]{tang2023neural}%
  \BibitemOpen
  \bibfield  {author} {\bibinfo {author} {\bibfnamefont {Y.}~\bibnamefont
  {Tang}}, \bibinfo {author} {\bibfnamefont {J.}~\bibnamefont {Weng}},\ and\
  \bibinfo {author} {\bibfnamefont {P.}~\bibnamefont {Zhang}},\ }\bibfield
  {title} {\bibinfo {title} {Neural-network solutions to stochastic reaction
  networks},\ }\href {https://www.nature.com/articles/s42256-023-00632-6}
  {\bibfield  {journal} {\bibinfo  {journal} {Nat. Mach. Intell.}\ }\textbf
  {\bibinfo {volume} {5}},\ \bibinfo {pages} {376} (\bibinfo {year}
  {2023})}\BibitemShut {NoStop}%
\bibitem [{\citenamefont {Forgoston}\ and\ \citenamefont
  {Moore}(2018)}]{forgoston2018primer}%
  \BibitemOpen
  \bibfield  {author} {\bibinfo {author} {\bibfnamefont {E.}~\bibnamefont
  {Forgoston}}\ and\ \bibinfo {author} {\bibfnamefont {R.~O.}\ \bibnamefont
  {Moore}},\ }\bibfield  {title} {\bibinfo {title} {A primer on noise-induced
  transitions in applied dynamical systems},\ }\href
  {https://doi.org/https://doi.org/10.1137/17M1142028} {\bibfield  {journal}
  {\bibinfo  {journal} {SIAM Rev.}\ }\textbf {\bibinfo {volume} {60}},\
  \bibinfo {pages} {969} (\bibinfo {year} {2018})}\BibitemShut {NoStop}%
\bibitem [{\citenamefont {Hartmann}\ \emph {et~al.}(2013)\citenamefont
  {Hartmann}, \citenamefont {Banisch}, \citenamefont {Sarich}, \citenamefont
  {Badowski},\ and\ \citenamefont
  {Sch{\"u}tte}}]{hartmann2013characterization}%
  \BibitemOpen
  \bibfield  {author} {\bibinfo {author} {\bibfnamefont {C.}~\bibnamefont
  {Hartmann}}, \bibinfo {author} {\bibfnamefont {R.}~\bibnamefont {Banisch}},
  \bibinfo {author} {\bibfnamefont {M.}~\bibnamefont {Sarich}}, \bibinfo
  {author} {\bibfnamefont {T.}~\bibnamefont {Badowski}},\ and\ \bibinfo
  {author} {\bibfnamefont {C.}~\bibnamefont {Sch{\"u}tte}},\ }\bibfield
  {title} {\bibinfo {title} {Characterization of rare events in molecular
  dynamics},\ }\href {https://doi.org/https://doi.org/10.3390/e16010350}
  {\bibfield  {journal} {\bibinfo  {journal} {Entropy}\ }\textbf {\bibinfo
  {volume} {16}},\ \bibinfo {pages} {350} (\bibinfo {year} {2013})}\BibitemShut
  {NoStop}%
\bibitem [{\citenamefont {Tanaka}\ \emph {et~al.}(2019)\citenamefont {Tanaka},
  \citenamefont {Yamane}, \citenamefont {H{\'e}roux}, \citenamefont {Nakane},
  \citenamefont {Kanazawa}, \citenamefont {Takeda}, \citenamefont {Numata},
  \citenamefont {Nakano},\ and\ \citenamefont {Hirose}}]{tanaka2019recent}%
  \BibitemOpen
  \bibfield  {author} {\bibinfo {author} {\bibfnamefont {G.}~\bibnamefont
  {Tanaka}}, \bibinfo {author} {\bibfnamefont {T.}~\bibnamefont {Yamane}},
  \bibinfo {author} {\bibfnamefont {J.~B.}\ \bibnamefont {H{\'e}roux}},
  \bibinfo {author} {\bibfnamefont {R.}~\bibnamefont {Nakane}}, \bibinfo
  {author} {\bibfnamefont {N.}~\bibnamefont {Kanazawa}}, \bibinfo {author}
  {\bibfnamefont {S.}~\bibnamefont {Takeda}}, \bibinfo {author} {\bibfnamefont
  {H.}~\bibnamefont {Numata}}, \bibinfo {author} {\bibfnamefont
  {D.}~\bibnamefont {Nakano}},\ and\ \bibinfo {author} {\bibfnamefont
  {A.}~\bibnamefont {Hirose}},\ }\bibfield  {title} {\bibinfo {title} {Recent
  advances in physical reservoir computing: A review},\ }\href
  {https://doi.org/https://doi.org/10.1016/j.neunet.2019.03.005} {\bibfield
  {journal} {\bibinfo  {journal} {Neural Netw.}\ }\textbf {\bibinfo {volume}
  {115}},\ \bibinfo {pages} {100} (\bibinfo {year} {2019})}\BibitemShut
  {NoStop}%
\bibitem [{\citenamefont {Xiong}\ and\ \citenamefont
  {Zhao}(2019)}]{xiong2019chaotic}%
  \BibitemOpen
  \bibfield  {author} {\bibinfo {author} {\bibfnamefont {Y.}~\bibnamefont
  {Xiong}}\ and\ \bibinfo {author} {\bibfnamefont {H.}~\bibnamefont {Zhao}},\
  }\bibfield  {title} {\bibinfo {title} {Chaotic time series prediction based
  on long short-term memory neural networks},\ }\href
  {https://doi.org/10.1360/SSPMA-2019-0115} {\bibfield  {journal} {\bibinfo
  {journal} {Sci. China Phy. Mech. Astron.}\ }\textbf {\bibinfo {volume}
  {49}},\ \bibinfo {pages} {120501} (\bibinfo {year} {2019})}\BibitemShut
  {NoStop}%
\bibitem [{\citenamefont {Karniadakis}\ \emph {et~al.}(2021)\citenamefont
  {Karniadakis}, \citenamefont {Kevrekidis}, \citenamefont {Lu}, \citenamefont
  {Perdikaris}, \citenamefont {Wang},\ and\ \citenamefont
  {Yang}}]{karniadakis2021physics}%
  \BibitemOpen
  \bibfield  {author} {\bibinfo {author} {\bibfnamefont {G.~E.}\ \bibnamefont
  {Karniadakis}}, \bibinfo {author} {\bibfnamefont {I.~G.}\ \bibnamefont
  {Kevrekidis}}, \bibinfo {author} {\bibfnamefont {L.}~\bibnamefont {Lu}},
  \bibinfo {author} {\bibfnamefont {P.}~\bibnamefont {Perdikaris}}, \bibinfo
  {author} {\bibfnamefont {S.}~\bibnamefont {Wang}},\ and\ \bibinfo {author}
  {\bibfnamefont {L.}~\bibnamefont {Yang}},\ }\bibfield  {title} {\bibinfo
  {title} {Physics-informed machine learning},\ }\href
  {https://doi.org/https://doi.org/10.1038/s42254-021-00314-5} {\bibfield
  {journal} {\bibinfo  {journal} {Nat. Rev. Phys.}\ }\textbf {\bibinfo {volume}
  {3}},\ \bibinfo {pages} {422} (\bibinfo {year} {2021})}\BibitemShut {NoStop}%
\bibitem [{\citenamefont {Zhao}(2021)}]{zhao2021inferring}%
  \BibitemOpen
  \bibfield  {author} {\bibinfo {author} {\bibfnamefont {H.}~\bibnamefont
  {Zhao}},\ }\bibfield  {title} {\bibinfo {title} {Inferring the dynamics of
  “black-box” systems using a learning machine},\ }\href
  {https://doi.org/https://doi.org/10.1007/s11433-021-1699-3} {\bibfield
  {journal} {\bibinfo  {journal} {Sci. China Phy. Mech. Astron.}\ }\textbf
  {\bibinfo {volume} {64}},\ \bibinfo {pages} {270511} (\bibinfo {year}
  {2021})}\BibitemShut {NoStop}%
\bibitem [{\citenamefont {Li}\ \emph {et~al.}(2023{\natexlab{a}})\citenamefont
  {Li}, \citenamefont {Wang}, \citenamefont {Roychowdhury},\ and\ \citenamefont
  {Jawed}}]{li2023meta}%
  \BibitemOpen
  \bibfield  {author} {\bibinfo {author} {\bibfnamefont {Q.}~\bibnamefont
  {Li}}, \bibinfo {author} {\bibfnamefont {T.}~\bibnamefont {Wang}}, \bibinfo
  {author} {\bibfnamefont {V.}~\bibnamefont {Roychowdhury}},\ and\ \bibinfo
  {author} {\bibfnamefont {M.~K.}\ \bibnamefont {Jawed}},\ }\bibfield  {title}
  {\bibinfo {title} {Meta-learning generalizable dynamics from trajectories},\
  }\href {https://doi.org/10.48550/arXiv.2301.00957} {\bibfield  {journal}
  {\bibinfo  {journal} {arXiv preprint arXiv:2301.00957}\ } (\bibinfo {year}
  {2023}{\natexlab{a}})}\BibitemShut {NoStop}%
\bibitem [{\citenamefont {Kaheman}\ \emph {et~al.}(2022)\citenamefont
  {Kaheman}, \citenamefont {Brunton},\ and\ \citenamefont
  {Kutz}}]{kaheman2022automatic}%
  \BibitemOpen
  \bibfield  {author} {\bibinfo {author} {\bibfnamefont {K.}~\bibnamefont
  {Kaheman}}, \bibinfo {author} {\bibfnamefont {S.~L.}\ \bibnamefont
  {Brunton}},\ and\ \bibinfo {author} {\bibfnamefont {J.~N.}\ \bibnamefont
  {Kutz}},\ }\bibfield  {title} {\bibinfo {title} {Automatic differentiation to
  simultaneously identify nonlinear dynamics and extract noise probability
  distributions from data},\ }\href {https://doi.org/10.1088/2632-2153/ac567a}
  {\bibfield  {journal} {\bibinfo  {journal} {Mach. Learn.: Sci. Technol.}\
  }\textbf {\bibinfo {volume} {3}},\ \bibinfo {pages} {015031} (\bibinfo {year}
  {2022})}\BibitemShut {NoStop}%
\bibitem [{\citenamefont {Raissi}\ \emph {et~al.}(2019)\citenamefont {Raissi},
  \citenamefont {Perdikaris},\ and\ \citenamefont
  {Karniadakis}}]{raissi2019physics}%
  \BibitemOpen
  \bibfield  {author} {\bibinfo {author} {\bibfnamefont {M.}~\bibnamefont
  {Raissi}}, \bibinfo {author} {\bibfnamefont {P.}~\bibnamefont {Perdikaris}},\
  and\ \bibinfo {author} {\bibfnamefont {G.~E.}\ \bibnamefont {Karniadakis}},\
  }\bibfield  {title} {\bibinfo {title} {Physics-informed neural networks: A
  deep learning framework for solving forward and inverse problems involving
  nonlinear partial differential equations},\ }\href
  {https://doi.org/https://doi.org/10.1016/j.jcp.2018.10.045} {\bibfield
  {journal} {\bibinfo  {journal} {J. Comput. Phys.}\ }\textbf {\bibinfo
  {volume} {378}},\ \bibinfo {pages} {686} (\bibinfo {year}
  {2019})}\BibitemShut {NoStop}%
\bibitem [{\citenamefont {Cuomo}\ \emph {et~al.}(2022)\citenamefont {Cuomo},
  \citenamefont {Di~Cola}, \citenamefont {Giampaolo}, \citenamefont {Rozza},
  \citenamefont {Raissi},\ and\ \citenamefont
  {Piccialli}}]{cuomo2022scientific}%
  \BibitemOpen
  \bibfield  {author} {\bibinfo {author} {\bibfnamefont {S.}~\bibnamefont
  {Cuomo}}, \bibinfo {author} {\bibfnamefont {V.~S.}\ \bibnamefont {Di~Cola}},
  \bibinfo {author} {\bibfnamefont {F.}~\bibnamefont {Giampaolo}}, \bibinfo
  {author} {\bibfnamefont {G.}~\bibnamefont {Rozza}}, \bibinfo {author}
  {\bibfnamefont {M.}~\bibnamefont {Raissi}},\ and\ \bibinfo {author}
  {\bibfnamefont {F.}~\bibnamefont {Piccialli}},\ }\bibfield  {title} {\bibinfo
  {title} {Scientific machine learning through physics--informed neural
  networks: Where we are and what’s next},\ }\href
  {https://doi.org/https://doi.org/10.1007/s10915-022-01939-z} {\bibfield
  {journal} {\bibinfo  {journal} {J. Sci. Comput.}\ }\textbf {\bibinfo {volume}
  {92}},\ \bibinfo {pages} {88} (\bibinfo {year} {2022})}\BibitemShut {NoStop}%
\bibitem [{\citenamefont {Lusch}\ \emph {et~al.}(2018)\citenamefont {Lusch},
  \citenamefont {Kutz},\ and\ \citenamefont {Brunton}}]{lusch2018deep}%
  \BibitemOpen
  \bibfield  {author} {\bibinfo {author} {\bibfnamefont {B.}~\bibnamefont
  {Lusch}}, \bibinfo {author} {\bibfnamefont {J.~N.}\ \bibnamefont {Kutz}},\
  and\ \bibinfo {author} {\bibfnamefont {S.~L.}\ \bibnamefont {Brunton}},\
  }\bibfield  {title} {\bibinfo {title} {Deep learning for universal linear
  embeddings of nonlinear dynamics},\ }\href
  {https://doi.org/https://doi.org/10.1038/s41467-018-07210-0} {\bibfield
  {journal} {\bibinfo  {journal} {Nat. Commun.}\ }\textbf {\bibinfo {volume}
  {9}},\ \bibinfo {pages} {4950} (\bibinfo {year} {2018})}\BibitemShut
  {NoStop}%
\bibitem [{\citenamefont {Jaeger}(2001)}]{jaeger2001echo}%
  \BibitemOpen
  \bibfield  {author} {\bibinfo {author} {\bibfnamefont {H.}~\bibnamefont
  {Jaeger}},\ }\bibfield  {title} {\bibinfo {title} {The “echo state”
  approach to analysing and training recurrent neural networks-with an erratum
  note},\ }\href@noop {} {\bibfield  {journal} {\bibinfo  {journal} {Bonn,
  Germany: German National Research Center for Information Technology GMD
  Technical Report}\ }\textbf {\bibinfo {volume} {148}},\ \bibinfo {pages} {13}
  (\bibinfo {year} {2001})}\BibitemShut {NoStop}%
\bibitem [{\citenamefont {Maass}\ \emph {et~al.}(2002)\citenamefont {Maass},
  \citenamefont {Natschl{\"a}ger},\ and\ \citenamefont
  {Markram}}]{maass2002real}%
  \BibitemOpen
  \bibfield  {author} {\bibinfo {author} {\bibfnamefont {W.}~\bibnamefont
  {Maass}}, \bibinfo {author} {\bibfnamefont {T.}~\bibnamefont
  {Natschl{\"a}ger}},\ and\ \bibinfo {author} {\bibfnamefont {H.}~\bibnamefont
  {Markram}},\ }\bibfield  {title} {\bibinfo {title} {Real-time computing
  without stable states: A new framework for neural computation based on
  perturbations},\ }\href {https://doi.org/10.1162/089976602760407955}
  {\bibfield  {journal} {\bibinfo  {journal} {Neural comput.}\ }\textbf
  {\bibinfo {volume} {14}},\ \bibinfo {pages} {2531} (\bibinfo {year}
  {2002})}\BibitemShut {NoStop}%
\bibitem [{\citenamefont {Pathak}\ \emph {et~al.}(2018)\citenamefont {Pathak},
  \citenamefont {Hunt}, \citenamefont {Girvan}, \citenamefont {Lu},\ and\
  \citenamefont {Ott}}]{pathak2018model}%
  \BibitemOpen
  \bibfield  {author} {\bibinfo {author} {\bibfnamefont {J.}~\bibnamefont
  {Pathak}}, \bibinfo {author} {\bibfnamefont {B.}~\bibnamefont {Hunt}},
  \bibinfo {author} {\bibfnamefont {M.}~\bibnamefont {Girvan}}, \bibinfo
  {author} {\bibfnamefont {Z.}~\bibnamefont {Lu}},\ and\ \bibinfo {author}
  {\bibfnamefont {E.}~\bibnamefont {Ott}},\ }\bibfield  {title} {\bibinfo
  {title} {Model-free prediction of large spatiotemporally chaotic systems from
  data: A reservoir computing approach},\ }\href
  {https://doi.org/https://doi.org/10.1103/PhysRevLett.120.024102} {\bibfield
  {journal} {\bibinfo  {journal} {Phys. Rev. Lett.}\ }\textbf {\bibinfo
  {volume} {120}},\ \bibinfo {pages} {024102} (\bibinfo {year}
  {2018})}\BibitemShut {NoStop}%
\bibitem [{\citenamefont {Kim}\ and\ \citenamefont
  {Bassett}(2023)}]{kim2023neural}%
  \BibitemOpen
  \bibfield  {author} {\bibinfo {author} {\bibfnamefont {J.~Z.}\ \bibnamefont
  {Kim}}\ and\ \bibinfo {author} {\bibfnamefont {D.~S.}\ \bibnamefont
  {Bassett}},\ }\bibfield  {title} {\bibinfo {title} {A neural machine code and
  programming framework for the reservoir computer},\ }\href
  {https://doi.org/10.1038/s42256-023-00668-8} {\bibfield  {journal} {\bibinfo
  {journal} {Nat. Mach. Intell.}\ }\textbf {\bibinfo {volume} {5}},\ \bibinfo
  {pages} {622} (\bibinfo {year} {2023})}\BibitemShut {NoStop}%
\bibitem [{\citenamefont {Jaeger}\ and\ \citenamefont
  {Haas}(2004)}]{jaeger2004harnessing}%
  \BibitemOpen
  \bibfield  {author} {\bibinfo {author} {\bibfnamefont {H.}~\bibnamefont
  {Jaeger}}\ and\ \bibinfo {author} {\bibfnamefont {H.}~\bibnamefont {Haas}},\
  }\bibfield  {title} {\bibinfo {title} {Harnessing nonlinearity: Predicting
  chaotic systems and saving energy in wireless communication},\ }\href
  {https://science.sciencemag.org/content/304/5667/78} {\bibfield  {journal}
  {\bibinfo  {journal} {Science}\ }\textbf {\bibinfo {volume} {304}},\ \bibinfo
  {pages} {78} (\bibinfo {year} {2004})}\BibitemShut {NoStop}%
\bibitem [{\citenamefont {Zimmermann}\ and\ \citenamefont
  {Parlitz}(2018)}]{zimmermann2018observing}%
  \BibitemOpen
  \bibfield  {author} {\bibinfo {author} {\bibfnamefont {R.~S.}\ \bibnamefont
  {Zimmermann}}\ and\ \bibinfo {author} {\bibfnamefont {U.}~\bibnamefont
  {Parlitz}},\ }\bibfield  {title} {\bibinfo {title} {Observing spatio-temporal
  dynamics of excitable media using reservoir computing},\ }\href
  {https://doi.org/https://doi.org/10.1063/1.5022276} {\bibfield  {journal}
  {\bibinfo  {journal} {Chaos}\ }\textbf {\bibinfo {volume} {28}},\ \bibinfo
  {pages} {043118} (\bibinfo {year} {2018})}\BibitemShut {NoStop}%
\bibitem [{\citenamefont {Fan}\ \emph {et~al.}(2020)\citenamefont {Fan},
  \citenamefont {Jiang}, \citenamefont {Zhang}, \citenamefont {Wang},\ and\
  \citenamefont {Lai}}]{fan2020long}%
  \BibitemOpen
  \bibfield  {author} {\bibinfo {author} {\bibfnamefont {H.}~\bibnamefont
  {Fan}}, \bibinfo {author} {\bibfnamefont {J.}~\bibnamefont {Jiang}}, \bibinfo
  {author} {\bibfnamefont {C.}~\bibnamefont {Zhang}}, \bibinfo {author}
  {\bibfnamefont {X.}~\bibnamefont {Wang}},\ and\ \bibinfo {author}
  {\bibfnamefont {Y.-C.}\ \bibnamefont {Lai}},\ }\bibfield  {title} {\bibinfo
  {title} {Long-term prediction of chaotic systems with machine learning},\
  }\href {https://doi.org/https://doi.org/10.1103/PhysRevResearch.2.012080}
  {\bibfield  {journal} {\bibinfo  {journal} {Phys. Rev. Res.}\ }\textbf
  {\bibinfo {volume} {2}},\ \bibinfo {pages} {012080} (\bibinfo {year}
  {2020})}\BibitemShut {NoStop}%
\bibitem [{\citenamefont {Kim}\ \emph {et~al.}(2021)\citenamefont {Kim},
  \citenamefont {Lu}, \citenamefont {Nozari}, \citenamefont {Pappas},\ and\
  \citenamefont {Bassett}}]{kim2021teaching}%
  \BibitemOpen
  \bibfield  {author} {\bibinfo {author} {\bibfnamefont {J.~Z.}\ \bibnamefont
  {Kim}}, \bibinfo {author} {\bibfnamefont {Z.}~\bibnamefont {Lu}}, \bibinfo
  {author} {\bibfnamefont {E.}~\bibnamefont {Nozari}}, \bibinfo {author}
  {\bibfnamefont {G.~J.}\ \bibnamefont {Pappas}},\ and\ \bibinfo {author}
  {\bibfnamefont {D.~S.}\ \bibnamefont {Bassett}},\ }\bibfield  {title}
  {\bibinfo {title} {Teaching recurrent neural networks to infer global
  temporal structure from local examples},\ }\href
  {https://doi.org/https://doi.org/10.1038/s42256-021-00321-2} {\bibfield
  {journal} {\bibinfo  {journal} {Nat. Mach. Intell.}\ }\textbf {\bibinfo
  {volume} {3}},\ \bibinfo {pages} {316} (\bibinfo {year} {2021})}\BibitemShut
  {NoStop}%
\bibitem [{\citenamefont {Zhai}\ \emph {et~al.}(2023)\citenamefont {Zhai},
  \citenamefont {Kong},\ and\ \citenamefont {Lai}}]{zhai2023emergence}%
  \BibitemOpen
  \bibfield  {author} {\bibinfo {author} {\bibfnamefont {Z.-M.}\ \bibnamefont
  {Zhai}}, \bibinfo {author} {\bibfnamefont {L.-W.}\ \bibnamefont {Kong}},\
  and\ \bibinfo {author} {\bibfnamefont {Y.-C.}\ \bibnamefont {Lai}},\
  }\bibfield  {title} {\bibinfo {title} {Emergence of a resonance in machine
  learning},\ }\href
  {https://doi.org/https://doi.org/10.1103/PhysRevResearch.5.033127} {\bibfield
   {journal} {\bibinfo  {journal} {Phys. Rev. Res.}\ }\textbf {\bibinfo
  {volume} {5}},\ \bibinfo {pages} {033127} (\bibinfo {year}
  {2023})}\BibitemShut {NoStop}%
\bibitem [{\citenamefont {Lim}\ \emph {et~al.}(2020)\citenamefont {Lim},
  \citenamefont {Theo~Giorgini}, \citenamefont {Moon},\ and\ \citenamefont
  {Wettlaufer}}]{Lim2020Predicting}%
  \BibitemOpen
  \bibfield  {author} {\bibinfo {author} {\bibfnamefont {S.~H.}\ \bibnamefont
  {Lim}}, \bibinfo {author} {\bibfnamefont {L.}~\bibnamefont {Theo~Giorgini}},
  \bibinfo {author} {\bibfnamefont {W.}~\bibnamefont {Moon}},\ and\ \bibinfo
  {author} {\bibfnamefont {J.~S.}\ \bibnamefont {Wettlaufer}},\ }\bibfield
  {title} {\bibinfo {title} {Predicting critical transitions in multiscale
  dynamical systems using reservoir computing},\ }\href
  {https://doi.org/10.1063/5.0023764} {\bibfield  {journal} {\bibinfo
  {journal} {Chaos}\ }\textbf {\bibinfo {volume} {30}},\ \bibinfo {pages}
  {123126} (\bibinfo {year} {2020})}\BibitemShut {NoStop}%
\bibitem [{\citenamefont {Carroll}(2018)}]{carroll2018using}%
  \BibitemOpen
  \bibfield  {author} {\bibinfo {author} {\bibfnamefont {T.~L.}\ \bibnamefont
  {Carroll}},\ }\bibfield  {title} {\bibinfo {title} {Using reservoir computers
  to distinguish chaotic signals},\ }\href
  {https://doi.org/https://doi.org/10.1103/PhysRevE.98.052209} {\bibfield
  {journal} {\bibinfo  {journal} {Phys. Rev. E}\ }\textbf {\bibinfo {volume}
  {98}},\ \bibinfo {pages} {052209} (\bibinfo {year} {2018})}\BibitemShut
  {NoStop}%
\bibitem [{\citenamefont {Nakai}\ and\ \citenamefont
  {Saiki}(2018)}]{nakai2018machine}%
  \BibitemOpen
  \bibfield  {author} {\bibinfo {author} {\bibfnamefont {K.}~\bibnamefont
  {Nakai}}\ and\ \bibinfo {author} {\bibfnamefont {Y.}~\bibnamefont {Saiki}},\
  }\bibfield  {title} {\bibinfo {title} {Machine-learning inference of fluid
  variables from data using reservoir computing},\ }\href
  {https://doi.org/https://doi.org/10.1103/PhysRevE.98.023111} {\bibfield
  {journal} {\bibinfo  {journal} {Phys. Rev. E}\ }\textbf {\bibinfo {volume}
  {98}},\ \bibinfo {pages} {023111} (\bibinfo {year} {2018})}\BibitemShut
  {NoStop}%
\bibitem [{\citenamefont {Weng}\ \emph {et~al.}(2019)\citenamefont {Weng},
  \citenamefont {Yang}, \citenamefont {Gu}, \citenamefont {Zhang},\ and\
  \citenamefont {Small}}]{weng2019synchronization}%
  \BibitemOpen
  \bibfield  {author} {\bibinfo {author} {\bibfnamefont {T.}~\bibnamefont
  {Weng}}, \bibinfo {author} {\bibfnamefont {H.}~\bibnamefont {Yang}}, \bibinfo
  {author} {\bibfnamefont {C.}~\bibnamefont {Gu}}, \bibinfo {author}
  {\bibfnamefont {J.}~\bibnamefont {Zhang}},\ and\ \bibinfo {author}
  {\bibfnamefont {M.}~\bibnamefont {Small}},\ }\bibfield  {title} {\bibinfo
  {title} {Synchronization of chaotic systems and their machine-learning
  models},\ }\href {https://doi.org/https://doi.org/10.1103/PhysRevE.99.042203}
  {\bibfield  {journal} {\bibinfo  {journal} {Phys. Rev. E}\ }\textbf {\bibinfo
  {volume} {99}},\ \bibinfo {pages} {042203} (\bibinfo {year}
  {2019})}\BibitemShut {NoStop}%
\bibitem [{\citenamefont {Grigoryeva}\ and\ \citenamefont
  {Ortega}(2018)}]{grigoryeva2018echo}%
  \BibitemOpen
  \bibfield  {author} {\bibinfo {author} {\bibfnamefont {L.}~\bibnamefont
  {Grigoryeva}}\ and\ \bibinfo {author} {\bibfnamefont {J.-P.}\ \bibnamefont
  {Ortega}},\ }\bibfield  {title} {\bibinfo {title} {Echo state networks are
  universal},\ }\href
  {https://doi.org/https://doi.org/10.1016/j.neunet.2018.08.025} {\bibfield
  {journal} {\bibinfo  {journal} {Neural Netw.}\ }\textbf {\bibinfo {volume}
  {108}},\ \bibinfo {pages} {495} (\bibinfo {year} {2018})}\BibitemShut
  {NoStop}%
\bibitem [{\citenamefont {Zhang}\ \emph {et~al.}(2021)\citenamefont {Zhang},
  \citenamefont {Fan}, \citenamefont {Wang},\ and\ \citenamefont
  {Wang}}]{zhang2021learning}%
  \BibitemOpen
  \bibfield  {author} {\bibinfo {author} {\bibfnamefont {H.}~\bibnamefont
  {Zhang}}, \bibinfo {author} {\bibfnamefont {H.}~\bibnamefont {Fan}}, \bibinfo
  {author} {\bibfnamefont {L.}~\bibnamefont {Wang}},\ and\ \bibinfo {author}
  {\bibfnamefont {X.}~\bibnamefont {Wang}},\ }\bibfield  {title} {\bibinfo
  {title} {Learning hamiltonian dynamics with reservoir computing},\ }\href
  {https://doi.org/https://doi.org/10.1103/PhysRevE.104.024205} {\bibfield
  {journal} {\bibinfo  {journal} {Phys. Rev. E}\ }\textbf {\bibinfo {volume}
  {104}},\ \bibinfo {pages} {024205} (\bibinfo {year} {2021})}\BibitemShut
  {NoStop}%
\bibitem [{\citenamefont {Tanaka}\ \emph {et~al.}(2022)\citenamefont {Tanaka},
  \citenamefont {Matsumori}, \citenamefont {Yoshida},\ and\ \citenamefont
  {Aihara}}]{tanaka2021reservoir}%
  \BibitemOpen
  \bibfield  {author} {\bibinfo {author} {\bibfnamefont {G.}~\bibnamefont
  {Tanaka}}, \bibinfo {author} {\bibfnamefont {T.}~\bibnamefont {Matsumori}},
  \bibinfo {author} {\bibfnamefont {H.}~\bibnamefont {Yoshida}},\ and\ \bibinfo
  {author} {\bibfnamefont {K.}~\bibnamefont {Aihara}},\ }\bibfield  {title}
  {\bibinfo {title} {Reservoir computing with diverse timescales for prediction
  of multiscale dynamics},\ }\href
  {https://doi.org/https://doi.org/10.1103/PhysRevResearch.4.L032014}
  {\bibfield  {journal} {\bibinfo  {journal} {Phys. Rev. Res.}\ }\textbf
  {\bibinfo {volume} {4}},\ \bibinfo {pages} {L032014} (\bibinfo {year}
  {2022})}\BibitemShut {NoStop}%
\bibitem [{\citenamefont {Du}\ \emph {et~al.}(2023)\citenamefont {Du},
  \citenamefont {Li}, \citenamefont {Fan}, \citenamefont {Zhan}, \citenamefont
  {Xiao},\ and\ \citenamefont {Wang}}]{du2023inferring}%
  \BibitemOpen
  \bibfield  {author} {\bibinfo {author} {\bibfnamefont {Y.}~\bibnamefont
  {Du}}, \bibinfo {author} {\bibfnamefont {Q.}~\bibnamefont {Li}}, \bibinfo
  {author} {\bibfnamefont {H.}~\bibnamefont {Fan}}, \bibinfo {author}
  {\bibfnamefont {M.}~\bibnamefont {Zhan}}, \bibinfo {author} {\bibfnamefont
  {J.}~\bibnamefont {Xiao}},\ and\ \bibinfo {author} {\bibfnamefont
  {X.}~\bibnamefont {Wang}},\ }\bibfield  {title} {\bibinfo {title} {Inferring
  attracting basins of power system with machine learning},\ }\href
  {https://doi.org/10.48550/arXiv.2305.14374} {\bibfield  {journal} {\bibinfo
  {journal} {arXiv preprint arXiv:2305.14374}\ } (\bibinfo {year}
  {2023})}\BibitemShut {NoStop}%
\bibitem [{\citenamefont {Fang}\ \emph {et~al.}(2023)\citenamefont {Fang},
  \citenamefont {Lu}, \citenamefont {Gao},\ and\ \citenamefont
  {Duan}}]{fang2023reservoir}%
  \BibitemOpen
  \bibfield  {author} {\bibinfo {author} {\bibfnamefont {C.}~\bibnamefont
  {Fang}}, \bibinfo {author} {\bibfnamefont {Y.}~\bibnamefont {Lu}}, \bibinfo
  {author} {\bibfnamefont {T.}~\bibnamefont {Gao}},\ and\ \bibinfo {author}
  {\bibfnamefont {J.}~\bibnamefont {Duan}},\ }\bibfield  {title} {\bibinfo
  {title} {Reservoir computing with error correction: Long-term behaviors of
  stochastic dynamical systems},\ }\href
  {https://doi.org/10.48550/arXiv.2305.00669} {\bibfield  {journal} {\bibinfo
  {journal} {arXiv preprint arXiv:2305.00669}\ } (\bibinfo {year}
  {2023})}\BibitemShut {NoStop}%
\bibitem [{\citenamefont {Tang}\ \emph {et~al.}(2018)\citenamefont {Tang},
  \citenamefont {Xu},\ and\ \citenamefont {Ao}}]{tang2018escape}%
  \BibitemOpen
  \bibfield  {author} {\bibinfo {author} {\bibfnamefont {Y.}~\bibnamefont
  {Tang}}, \bibinfo {author} {\bibfnamefont {S.}~\bibnamefont {Xu}},\ and\
  \bibinfo {author} {\bibfnamefont {P.}~\bibnamefont {Ao}},\ }\bibfield
  {title} {\bibinfo {title} {Escape rate for nonequilibrium processes dominated
  by strong non-detailed balance force},\ }\href
  {https://doi.org/https://doi.org/10.1063/1.5008524} {\bibfield  {journal}
  {\bibinfo  {journal} {J. Chem. Phys.}\ }\textbf {\bibinfo {volume} {148}},\
  \bibinfo {pages} {064102} (\bibinfo {year} {2018})}\BibitemShut {NoStop}%
\bibitem [{\citenamefont {Belkacemi}\ \emph {et~al.}(2021)\citenamefont
  {Belkacemi}, \citenamefont {Gkeka}, \citenamefont {Leli{\`e}vre},\ and\
  \citenamefont {Stoltz}}]{belkacemi2021chasing}%
  \BibitemOpen
  \bibfield  {author} {\bibinfo {author} {\bibfnamefont {Z.}~\bibnamefont
  {Belkacemi}}, \bibinfo {author} {\bibfnamefont {P.}~\bibnamefont {Gkeka}},
  \bibinfo {author} {\bibfnamefont {T.}~\bibnamefont {Leli{\`e}vre}},\ and\
  \bibinfo {author} {\bibfnamefont {G.}~\bibnamefont {Stoltz}},\ }\bibfield
  {title} {\bibinfo {title} {Chasing collective variables using autoencoders
  and biased trajectories},\ }\href {https://doi.org/10.1021/acs.jctc.1c00415}
  {\bibfield  {journal} {\bibinfo  {journal} {J. Chem. Theory Comput.}\
  }\textbf {\bibinfo {volume} {18}},\ \bibinfo {pages} {59} (\bibinfo {year}
  {2021})}\BibitemShut {NoStop}%
\bibitem [{\citenamefont
  {Luko{\v{s}}evi{\v{c}}ius}(2012)}]{lukovsevivcius2012practical}%
  \BibitemOpen
  \bibfield  {author} {\bibinfo {author} {\bibfnamefont {M.}~\bibnamefont
  {Luko{\v{s}}evi{\v{c}}ius}},\ }\bibfield  {title} {\bibinfo {title} {A
  practical guide to applying echo state networks},\ }in\ \href@noop {} {\emph
  {\bibinfo {booktitle} {Neural Networks: Tricks of the Trade: Second
  Edition}}}\ (\bibinfo  {publisher} {Springer},\ \bibinfo {year} {2012})\ pp.\
  \bibinfo {pages} {659--686}\BibitemShut {NoStop}%
\bibitem [{\citenamefont {Lorenz}(1963)}]{lorenz1963deterministic}%
  \BibitemOpen
  \bibfield  {author} {\bibinfo {author} {\bibfnamefont {E.~N.}\ \bibnamefont
  {Lorenz}},\ }\bibfield  {title} {\bibinfo {title} {Deterministic nonperiodic
  flow},\ }\href
  {https://doi.org/https://doi.org/10.1175/1520-0469(1963)020<0130:DNF>2.0.CO;2}
  {\bibfield  {journal} {\bibinfo  {journal} {J. Atmos. Sci.}\ }\textbf
  {\bibinfo {volume} {20}},\ \bibinfo {pages} {130} (\bibinfo {year}
  {1963})}\BibitemShut {NoStop}%
\bibitem [{\citenamefont {Jiang}\ and\ \citenamefont
  {Lai}(2019)}]{PhysRevResearch.1.033056}%
  \BibitemOpen
  \bibfield  {author} {\bibinfo {author} {\bibfnamefont {J.}~\bibnamefont
  {Jiang}}\ and\ \bibinfo {author} {\bibfnamefont {Y.-C.}\ \bibnamefont
  {Lai}},\ }\bibfield  {title} {\bibinfo {title} {Model-free prediction of
  spatiotemporal dynamical systems with recurrent neural networks: Role of
  network spectral radius},\ }\href
  {https://doi.org/10.1103/PhysRevResearch.1.033056} {\bibfield  {journal}
  {\bibinfo  {journal} {Phys. Rev. Research}\ }\textbf {\bibinfo {volume}
  {1}},\ \bibinfo {pages} {033056} (\bibinfo {year} {2019})}\BibitemShut
  {NoStop}%
\bibitem [{\citenamefont {Gauthier}\ \emph {et~al.}(2021)\citenamefont
  {Gauthier}, \citenamefont {Bollt}, \citenamefont {Griffith},\ and\
  \citenamefont {Barbosa}}]{gauthier2021next}%
  \BibitemOpen
  \bibfield  {author} {\bibinfo {author} {\bibfnamefont {D.~J.}\ \bibnamefont
  {Gauthier}}, \bibinfo {author} {\bibfnamefont {E.}~\bibnamefont {Bollt}},
  \bibinfo {author} {\bibfnamefont {A.}~\bibnamefont {Griffith}},\ and\
  \bibinfo {author} {\bibfnamefont {W.~A.}\ \bibnamefont {Barbosa}},\
  }\bibfield  {title} {\bibinfo {title} {Next generation reservoir computing},\
  }\href {https://www.nature.com/articles/s41467-021-25801-2} {\bibfield
  {journal} {\bibinfo  {journal} {Nat. Commun.}\ }\textbf {\bibinfo {volume}
  {12}},\ \bibinfo {pages} {5564} (\bibinfo {year} {2021})}\BibitemShut
  {NoStop}%
\bibitem [{\citenamefont {Yperman}\ and\ \citenamefont
  {Becker}(2016)}]{yperman2016bayesian}%
  \BibitemOpen
  \bibfield  {author} {\bibinfo {author} {\bibfnamefont {J.}~\bibnamefont
  {Yperman}}\ and\ \bibinfo {author} {\bibfnamefont {T.}~\bibnamefont
  {Becker}},\ }\bibfield  {title} {\bibinfo {title} {Bayesian optimization of
  hyper-parameters in reservoir computing},\ }\href
  {https://doi.org/10.48550/arXiv.1611.05193} {\bibfield  {journal} {\bibinfo
  {journal} {arXiv preprint arXiv:1611.05193}\ } (\bibinfo {year}
  {2016})}\BibitemShut {NoStop}%
\bibitem [{\citenamefont {Ren}\ and\ \citenamefont {Ma}(2022)}]{ren2022global}%
  \BibitemOpen
  \bibfield  {author} {\bibinfo {author} {\bibfnamefont {B.}~\bibnamefont
  {Ren}}\ and\ \bibinfo {author} {\bibfnamefont {H.}~\bibnamefont {Ma}},\
  }\bibfield  {title} {\bibinfo {title} {Global optimization of
  hyper-parameters in reservoir computing},\ }\href
  {https://doi.org/10.3934/era.2022139} {\bibfield  {journal} {\bibinfo
  {journal} {Electron. Res. Arch.}\ }\textbf {\bibinfo {volume} {30}},\
  \bibinfo {pages} {2719} (\bibinfo {year} {2022})}\BibitemShut {NoStop}%
\bibitem [{\citenamefont {Hoffmann}\ \emph {et~al.}(2021)\citenamefont
  {Hoffmann}, \citenamefont {Scherer}, \citenamefont {Hempel}, \citenamefont
  {Mardt}, \citenamefont {de~Silva}, \citenamefont {Husic}, \citenamefont
  {Klus}, \citenamefont {Wu}, \citenamefont {Kutz}, \citenamefont {Brunton}
  \emph {et~al.}}]{hoffmann2021deeptime}%
  \BibitemOpen
  \bibfield  {author} {\bibinfo {author} {\bibfnamefont {M.}~\bibnamefont
  {Hoffmann}}, \bibinfo {author} {\bibfnamefont {M.}~\bibnamefont {Scherer}},
  \bibinfo {author} {\bibfnamefont {T.}~\bibnamefont {Hempel}}, \bibinfo
  {author} {\bibfnamefont {A.}~\bibnamefont {Mardt}}, \bibinfo {author}
  {\bibfnamefont {B.}~\bibnamefont {de~Silva}}, \bibinfo {author}
  {\bibfnamefont {B.~E.}\ \bibnamefont {Husic}}, \bibinfo {author}
  {\bibfnamefont {S.}~\bibnamefont {Klus}}, \bibinfo {author} {\bibfnamefont
  {H.}~\bibnamefont {Wu}}, \bibinfo {author} {\bibfnamefont {N.}~\bibnamefont
  {Kutz}}, \bibinfo {author} {\bibfnamefont {S.~L.}\ \bibnamefont {Brunton}},
  \emph {et~al.},\ }\bibfield  {title} {\bibinfo {title} {{Deeptime: a Python
  library for machine learning dynamical models from time series data}},\
  }\href {https://doi.org/10.1088/2632-2153/ac3de0} {\bibfield  {journal}
  {\bibinfo  {journal} {Mach. Learn.: Sci. Technol. 3 015009}\ }\textbf
  {\bibinfo {volume} {3}},\ \bibinfo {pages} {015009} (\bibinfo {year}
  {2021})}\BibitemShut {NoStop}%
\bibitem [{\citenamefont {Li}\ \emph {et~al.}(2023{\natexlab{b}})\citenamefont
  {Li}, \citenamefont {Wang},\ and\ \citenamefont {Li}}]{li2023learning}%
  \BibitemOpen
  \bibfield  {author} {\bibinfo {author} {\bibfnamefont {R.}~\bibnamefont
  {Li}}, \bibinfo {author} {\bibfnamefont {H.}~\bibnamefont {Wang}},\ and\
  \bibinfo {author} {\bibfnamefont {Y.}~\bibnamefont {Li}},\ }\bibfield
  {title} {\bibinfo {title} {Learning slow and fast system dynamics via
  automatic separation of time scales},\ }in\ \href
  {https://doi.org/10.1145/3580305.3599858} {\emph {\bibinfo {booktitle} {Proc.
  ACM SIGKDD Int. Conf. Knowl. Discov. Data Min.}}},\ \bibinfo {series and
  number} {KDD '23}\ (\bibinfo  {publisher} {Association for Computing
  Machinery},\ \bibinfo {address} {New York, NY, USA},\ \bibinfo {year}
  {2023})\ p.\ \bibinfo {pages} {4380–4390}\BibitemShut {NoStop}%
\bibitem [{\citenamefont {Casert}\ \emph {et~al.}(2022)\citenamefont {Casert},
  \citenamefont {Tamblyn},\ and\ \citenamefont
  {Whitelam}}]{casert2022learning}%
  \BibitemOpen
  \bibfield  {author} {\bibinfo {author} {\bibfnamefont {C.}~\bibnamefont
  {Casert}}, \bibinfo {author} {\bibfnamefont {I.}~\bibnamefont {Tamblyn}},\
  and\ \bibinfo {author} {\bibfnamefont {S.}~\bibnamefont {Whitelam}},\
  }\bibfield  {title} {\bibinfo {title} {Learning stochastic dynamics and
  predicting emergent behavior using transformers},\ }\href
  {https://arxiv.org/abs/2202.08708} {\bibfield  {journal} {\bibinfo  {journal}
  {arXiv preprint arXiv:2202.08708}\ } (\bibinfo {year} {2022})}\BibitemShut
  {NoStop}%
\bibitem [{\citenamefont {Tang}\ \emph {et~al.}(2022)\citenamefont {Tang},
  \citenamefont {Liu}, \citenamefont {Zhang},\ and\ \citenamefont
  {Zhang}}]{tang2022solving}%
  \BibitemOpen
  \bibfield  {author} {\bibinfo {author} {\bibfnamefont {Y.}~\bibnamefont
  {Tang}}, \bibinfo {author} {\bibfnamefont {J.}~\bibnamefont {Liu}}, \bibinfo
  {author} {\bibfnamefont {J.}~\bibnamefont {Zhang}},\ and\ \bibinfo {author}
  {\bibfnamefont {P.}~\bibnamefont {Zhang}},\ }\bibfield  {title} {\bibinfo
  {title} {Solving nonequilibrium statistical mechanics by evolving
  autoregressive neural networks},\ }\href {https://arxiv.org/abs/2208.08266}
  {\bibfield  {journal} {\bibinfo  {journal} {arXiv preprint arXiv:2208.08266}\
  } (\bibinfo {year} {2022})}\BibitemShut {NoStop}%
\bibitem [{\citenamefont {Vanden-Eijnden}\ and\ \citenamefont
  {Weare}(2013)}]{vanden2013data}%
  \BibitemOpen
  \bibfield  {author} {\bibinfo {author} {\bibfnamefont {E.}~\bibnamefont
  {Vanden-Eijnden}}\ and\ \bibinfo {author} {\bibfnamefont {J.}~\bibnamefont
  {Weare}},\ }\bibfield  {title} {\bibinfo {title} {Data assimilation in the
  low noise regime with application to the kuroshio},\ }\href
  {https://doi.org/https://doi.org/10.1175/MWR-D-12-00060.1} {\bibfield
  {journal} {\bibinfo  {journal} {Mon. Weather Rev.}\ }\textbf {\bibinfo
  {volume} {141}},\ \bibinfo {pages} {1822} (\bibinfo {year}
  {2013})}\BibitemShut {NoStop}%
\bibitem [{\citenamefont {Wunderlich}\ and\ \citenamefont
  {Sklar}(2022)}]{wunderlich2022data}%
  \BibitemOpen
  \bibfield  {author} {\bibinfo {author} {\bibfnamefont {A.}~\bibnamefont
  {Wunderlich}}\ and\ \bibinfo {author} {\bibfnamefont {J.}~\bibnamefont
  {Sklar}},\ }\bibfield  {title} {\bibinfo {title} {Data-driven modeling of
  noise time series with convolutional generative adversarial networks},\
  }\href {https://doi.org/10.1088/2632-2153/acee44} {\bibfield  {journal}
  {\bibinfo  {journal} {Mach. Learn.: Sci. Technol.}\ }\textbf {\bibinfo
  {volume} {4}},\ \bibinfo {pages} {035023} (\bibinfo {year}
  {2022})}\BibitemShut {NoStop}%
\bibitem [{\citenamefont {Verzelli}(2022)}]{verzelli2022learning}%
  \BibitemOpen
  \bibfield  {author} {\bibinfo {author} {\bibfnamefont {P.}~\bibnamefont
  {Verzelli}},\ }\emph {\bibinfo {title} {Learning dynamical systems using
  dynamical systems: the reservoir computing approach}},\ \href
  {https://sonar.ch/usi/documents/319318} {Ph.D. thesis},\ \bibinfo  {school}
  {Università della Svizzera italiana} (\bibinfo {year} {2022})\BibitemShut
  {NoStop}%
\end{thebibliography}%

\clearpage
\supplementarysection

\section{Supplementary note}
In the Supplementary Note, we first decrease the length of the training set to observe its effect on training. We also apply the present method to more systems, including a 1D tilted gradient system, a 2D gradient system, 2D tilted gradient and non-gradient systems, and a 2D tristable system. Figures of the results are in Supplementary Figures.

\subsection{Assessing the impact of training set length on performance} \label{fulu1}
To assess the performance of the present method with the limited training data, we shorten the length of the training set based on Example 1 of the main text. Additionally, we keep the length of the predicting set same as the training set. \autoref{differentLength}(a-b) display the same results in FIG.~2(e-f) of the main text, over the same hyperparameters and length ($10000\delta t$). Next, we reduce the length of the training and predicting sets by half ($5000\delta t$), as illustrated in \autoref{differentLength}(c-d). Without any significant changes in the results, we continue to reduce the length by half, this means that the training and predicting sets are only one-fourth of their original size ($2500\delta t$). However, the presented results (\autoref{differentLength}(e-f)) exhibit larger errors compared with the results in \autoref{differentLength}(a-d). Furthermore, we proceed to halve the length ($1250\delta t$) of both the training and predicting sets once again. The results depicted in \autoref{differentLength}(g-h) demonstrate significant errors. This example shows the training set length impact on our training.
\subsection{More examples}
\label{fulu2}
\subsubsection{A one-dimensional tilted bistable gradient system} \label{1D C section}
\label{Se1}

To demonstrate whether the noise-induced transitions can be predicted by the present method for systems with the tilted double-well potential, we apply the present method to the system depicted in \autoref{doublewell with c}. This system is the same as Example 1 of the main text, and the parameter $c=0.25$ makes the system to be tilted. The distinct time scales of upward and downward transitions necessitate two sets of hyperparameters for effective learning. The ``upward'' refers to the transitions from down state to up state, and the ``downward'' refers to the transitions from up state to down state. 

 \autoref{doublewell with c}(a) shows that we focus on the upward transitions, and \autoref{doublewell with c}(b-f) are the results. As the same method in Framework, we find proper hyperparameters set $1$ (for upward transitions) listed in \autoref{doublewell tilted hyperparameters} to obtain the trained slow-scale model (for upward transitions). The ten different slowly time-scale series and separated noise distribution as illustrated in \autoref{doublewell with c}(b). In the predicting phase, we perform rolling prediction with the trained slow-scale model (for upward transitions) and the noise distribution. Then, we evaluate the transitions within the test and predicted data.
 
  We note a significant difference in the downward transition time between test and predicted data (\autoref{doublewell with c}(c)), and with differences in the number of downward transitions per $10000\delta t$ (\autoref{doublewell with c}(d)). Meanwhile, our evaluation of the upward transition time (\autoref{doublewell with c}(e)) and the number of upward transitions in $10000\delta t$ (\autoref{doublewell with c}(f)) match. It indicates the effectiveness of hyperparameters set $1$ for upward transitions.
 
Using the same training set, we switch to downward transitions, with results shown in \autoref{doublewell with c}(g-l). As we shift focus, we observe the results similar to the upward case. With the proper hyperparameters set $2$ (in \autoref{doublewell tilted hyperparameters}) for downward transitions, we determine the corresponding slow-scale model (for downward transitions). The results, including downward transition time (\autoref{doublewell with c}(i)) and the number of downward transitions in $10000\delta t$ (\autoref{doublewell with c}(j)) match. Conversely, for upward transitions, a substantial error is observable in \autoref{doublewell with c}(k), while \autoref{doublewell with c}(l) shows the result with an insignificant error. It displays the effectiveness of hyperparameters set $2$ for downward transitions.

\subsubsection{A two-dimensional bistable gradient system}
\label{Se2}

Before the work in Example 3 of the main text, we design a pre-experiment, to ensure the present method can apply in a 2D bistable system. The system has noise-induced transitions between the two potential wells under the noise as illustrated in \autoref{2d doublewell a=0}(a), and the scatter plot of the training set (consisting of $20000$ data points) is depicted in \autoref{2d doublewell a=0}(b). According to the method in Framework, we generate ten different slowly time-scale series in the training phase (\autoref{2d doublewell a=0}(c)) and prediction (consisting of 20000 data points) in the predicting phase (\autoref{2d doublewell a=0}(d)). In \autoref{2d doublewell a=0}(e-f), we compare the number of transitions and the transition time (over $20000\delta t$) of generated data and predictions. The proper hyperparameters for this system are listed in \autoref{2d doublewell a0 hyperparameters}.

\subsubsection{A two-dimensional tilted bistable gradient system}
\label{Se3}
To demonstrate the present method can predict noise-induced transitions in a 2D tilted system, we first apply it to a 2D bistable gradient system with white noise as illustrated in \autoref{2d doublewell with c}.

Similar to the 1D tilted double-well potential system, the upward and the downward transitions necessitate two sets of hyperparameters. The ``upward'' refers to the transitions from down state to up state, and the ``downward'' refers
the transitions from up state to down state. As shown in \autoref{2d doublewell with c}, the same graphical scheme used in \autoref{doublewell with c} is applied to represent the results of the two-dimensional tilted bistable gradient system with Gaussian white noise. \autoref{2d doublewell with c}(a-f) show the results for upward transitions, and the hyperparameters set $1$ (for upward transitions) are listed in \autoref{2d doublewell tilted  hyperparameters}. \autoref{2d doublewell with c}(g-l) show the results for downward transitions, and the hyperparameters set $2$ (for downward transitions) are listed in \autoref{2d doublewell tilted  hyperparameters}.

\subsubsection{A two-dimensional tilted bistable non-gradient system}
\label{Se4}

To demonstrate the present method can predict noise-induced transitions in a both tilted and non-gradient system, we add a rotation element to make this system without detailed balance as illustrated in \autoref{2d doublewell with a+c} to apply our approach.

 As shown in \autoref{2d doublewell with a+c}, the same graphical scheme used in \autoref{2d doublewell with c} is applied to represent the results of a two-dimensional non-gradient system with double-well potential and Gaussian white noise. The ``upward'' refers to the transitions from down state to up state, and the ``downward'' refers
the transitions from up state to down state. \autoref{2d doublewell with a+c}(a-f) show the results for upward transitions, and the hyperparameters set $3$ (for upward transitions) are listed in \autoref{2d doublewell tilted  hyperparameters}. \autoref{2d doublewell with a+c}(g-l) show the results for downward transitions, and the hyperparameters set $4$ (for downward transitions) are listed in \autoref{2d doublewell tilted  hyperparameters}.

\subsubsection{A two-dimensional tristable system}\label{Se5}

 To assess the performance of the present method in multi-stable systems, we study a 2D tristable system \cite{belkacemi2021chasing} as illustrated in \autoref{2D-triple fig}(a):
\begin{align}
\begin{split}
    \dot{u}_1 &=10u_1e^{-u_1^2}\left[e^{-(u_2-1/3)^2}-e^{-(u_2-5/3)^2} \right]-8e^{-u_2^2}\left[ (u_1-1)e^{-(u_1-1)^2}+(u_1+1)e^{-(u_1+1)^2} \right] \\
    &\ -0.8u_1^3 + \sqrt{\frac{2}{\gamma}}\xi_1(t),\enspace t \ge 0,
\end{split}
\label{2-D triple in x}
\\
\begin{split}
    \dot{u}_2 &=-8u_2e^{-u_2^2}\left[ e^{-(u_1-1)^2}+ e^{-(u_1+1)^2} \right] +10e^{-u_1^2}\left[(u_2-\tfrac{1}{3})e^{-(u_2-1/3)^2}-(u_2-\tfrac{5}{3})e^{-(u_2-5/3)^2} \right]\\
    &\quad -0.8(u_2-\tfrac{1}{3})^3+\sqrt{\frac{2}{\gamma}}\xi_2(t),\enspace t \ge 0.
\end{split}
\label{2-D triple in y}
\end{align}The system has noise-induced transitions under the Gaussian white noise ($\xi_1(t), \xi_2(t)$) as in Eq.~(8) of the main text, and $\gamma$ corresponds to the noise strength. 

To investigate transitions among three stable states, we specifically consider the $u_{1}$ direction due to the overlapping of two wells in the $u_{2}$ direction. Therefore, our study focuses on a one-dimensional time series in the $u_{1}$ direction. We generated a time series from \cref{2-D triple in x,2-D triple in y,}, in $t\in[0,100]$. The time series is plotted for the first $30000\delta t$ in \autoref{2D-triple fig}(b).

In the training phase, with the training set $t\in[0,50]$, we obtain the appropriate hyperparameters listed in \autoref{Triple Hyperparameters}. The time series from ten distinct initial points converge to the stable states $(u_1=-1, u_1=0, u_1=1)$ as illustrated in \autoref{2D-triple fig}(c). Subsequently, we perform rolling prediction for $t\in [50,80]$ using the separated noise and the trained slow-scale model.  
The result demonstrates that the method learns the multi-state dynamics of the generated time series, including the transitions among these stable states. 



\clearpage
\section{Supplement figures}

\begin{figure}[!htbp]
    \centering
    \includegraphics[width=0.7\textwidth]{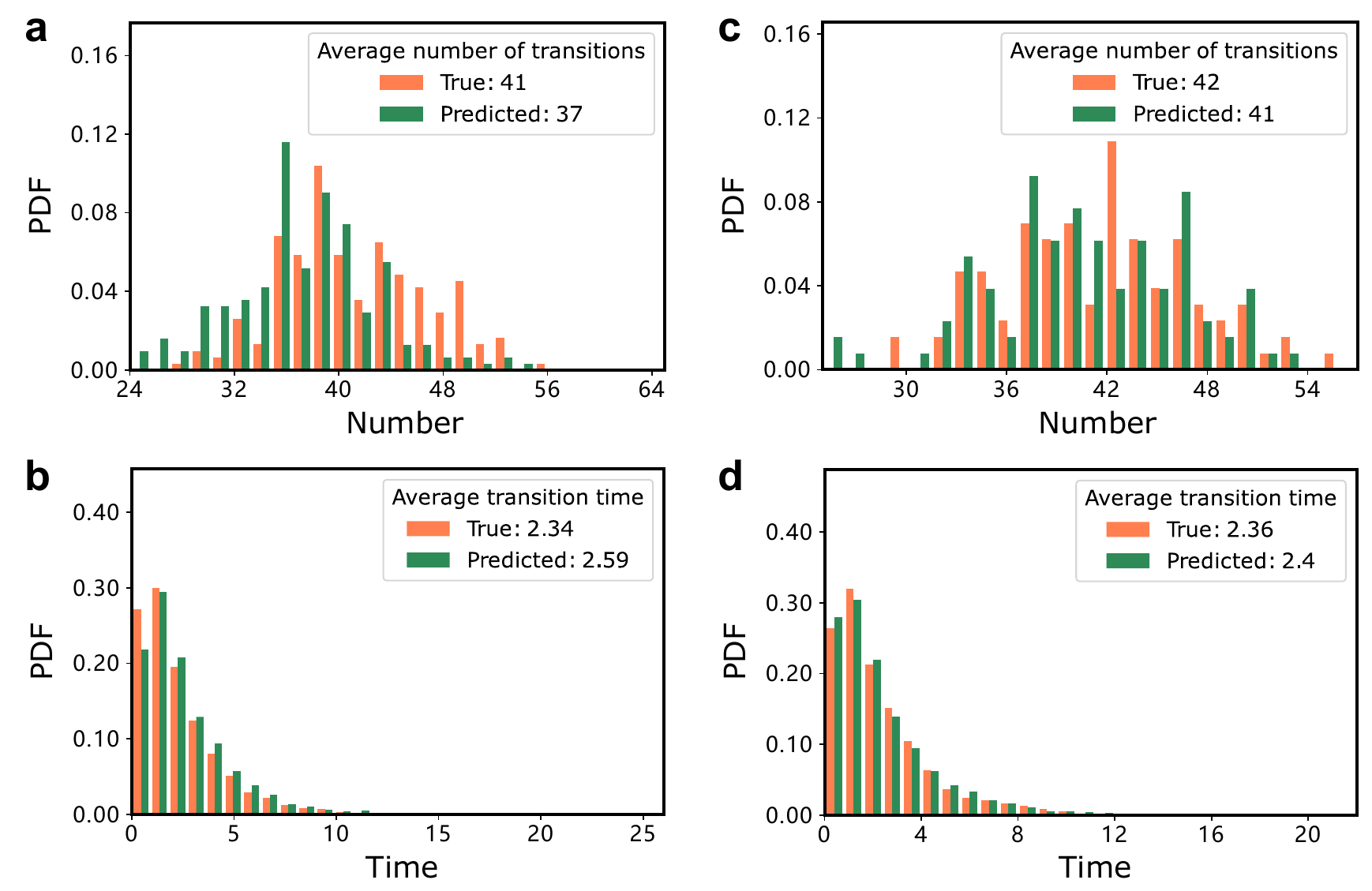}
    \caption{\textbf{Results of increasing the intensity of sampled noise in rolling prediction based on Example 1 of the main text.} PDF: probability density function. (a) The number of transitions of test and predicted data over $10000\delta t$. (b) Histograms of transition time of test and predicted data within $10000\delta t$. (c-d) Same as (a-b), and sampled noise is amplified by a factor of 1.1 in rolling prediction.}
    \label{noisefactor}
\end{figure}

\clearpage
\begin{figure}[!htbp]
    \centering
    \includegraphics[width=\textwidth]{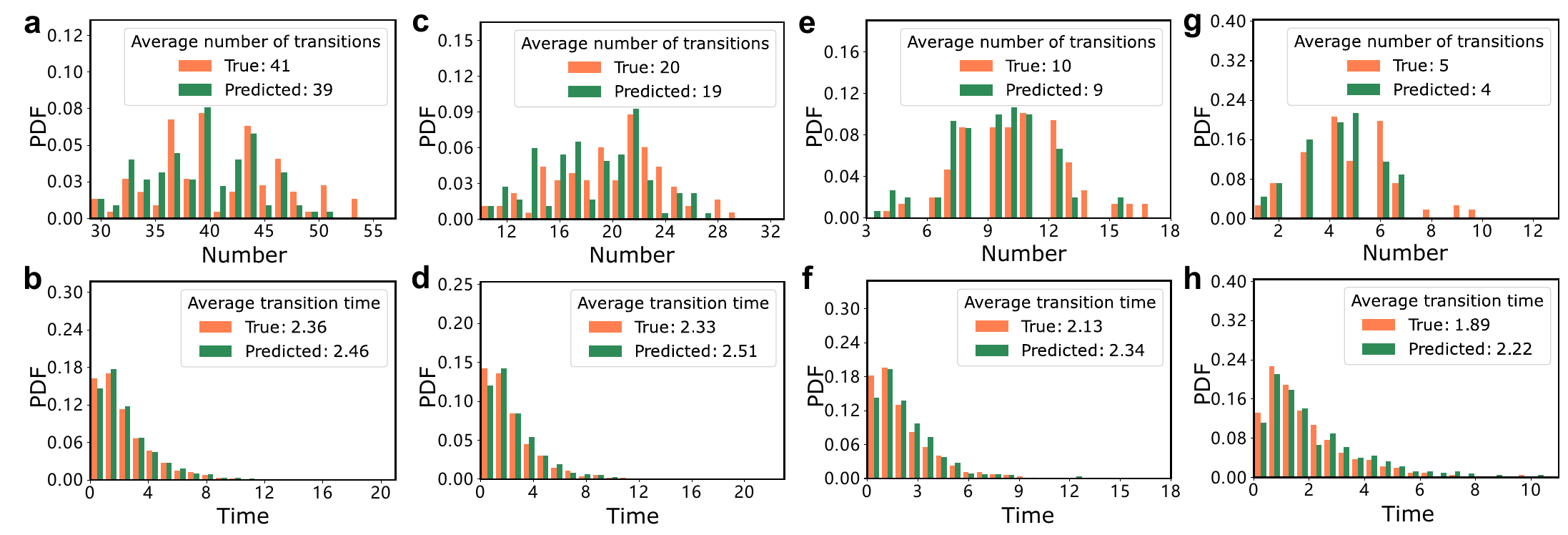}
    \caption{\textbf{The influence of decreasing the length of training and predicting sets based on Example 1 of the main text.} PDF: probability density function. (a) The number of transitions over the 10000 test and predicted data points. (b) Histograms of transition time of the 10000 test and predicted data points. (c-d) Results over the 5000 test and predicted data points. (e-f) Results over the 2500 test and predicted data points. (g-h) Results over the 1250 test and predicted data points.}
    \label{differentLength}
\end{figure}

\clearpage
\begin{figure}[!htbp]
    \centering
    \includegraphics[width=\textwidth]{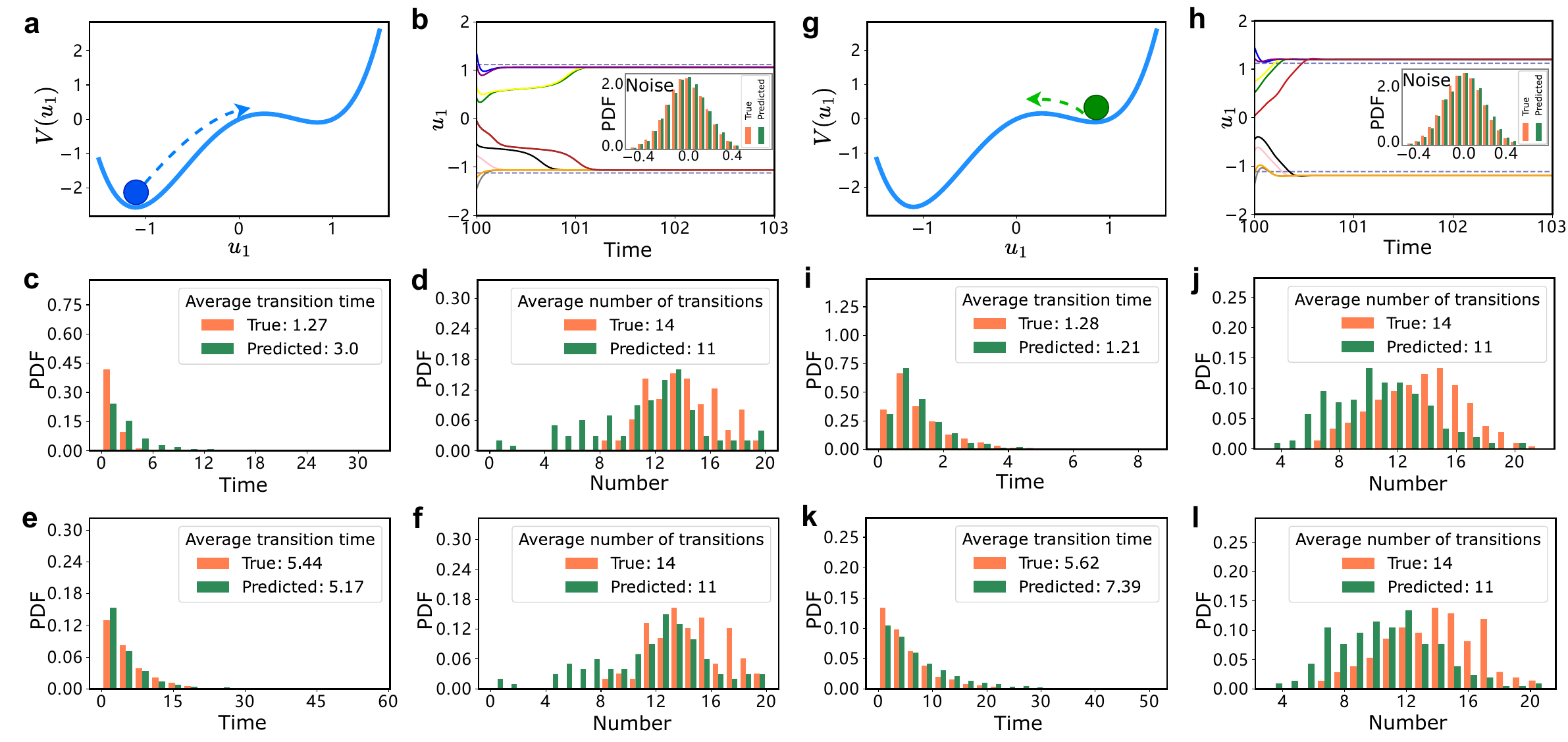}
    \caption{ \textbf{Capturing stochastic transitions in a 1D tilted bistable gradient system with white noise.} The system is the same as Example 1 of the main text, and the parameter $c=0.25$, making the potential tilted. Generated time series from Eq.~(9) of the main text ($b=5, c=0.25, \varepsilon=0.3, \delta t =0.01$), spanning a duration of $20000\delta t$, with the training
set $t\in[0,100]$ and the predicting set $t\in [100,200]$. (a) Schematic of noise-induced upward transitions in the 1D tilted bistable gradient system with Gaussian white noise. For figures (b-f), we focus on the upward transitions. (b) The trained slow-scale model transforms ten different start points into ten different slowly time-scale series (color lines). Separate the noise distribution. 
 (c) Histograms of downward transition time for the test and predicted data. (d) The number of downward transitions for the test and predicted data. The duration of the predicting set is $10000\delta t$ (10000 data points). (e) Histograms of upward transition time for the test and predicted data. (f) The number of upward transitions for the test and predicted data. The duration of the predicting set is $10000\delta t$ (10000 data points). (g-l) Same as (a-f), and focusing on the downward transitions. }
    \label{doublewell with c}
\end{figure}

\clearpage
\begin{figure}[!htbp]
    \centering
    \includegraphics[width=\textwidth]{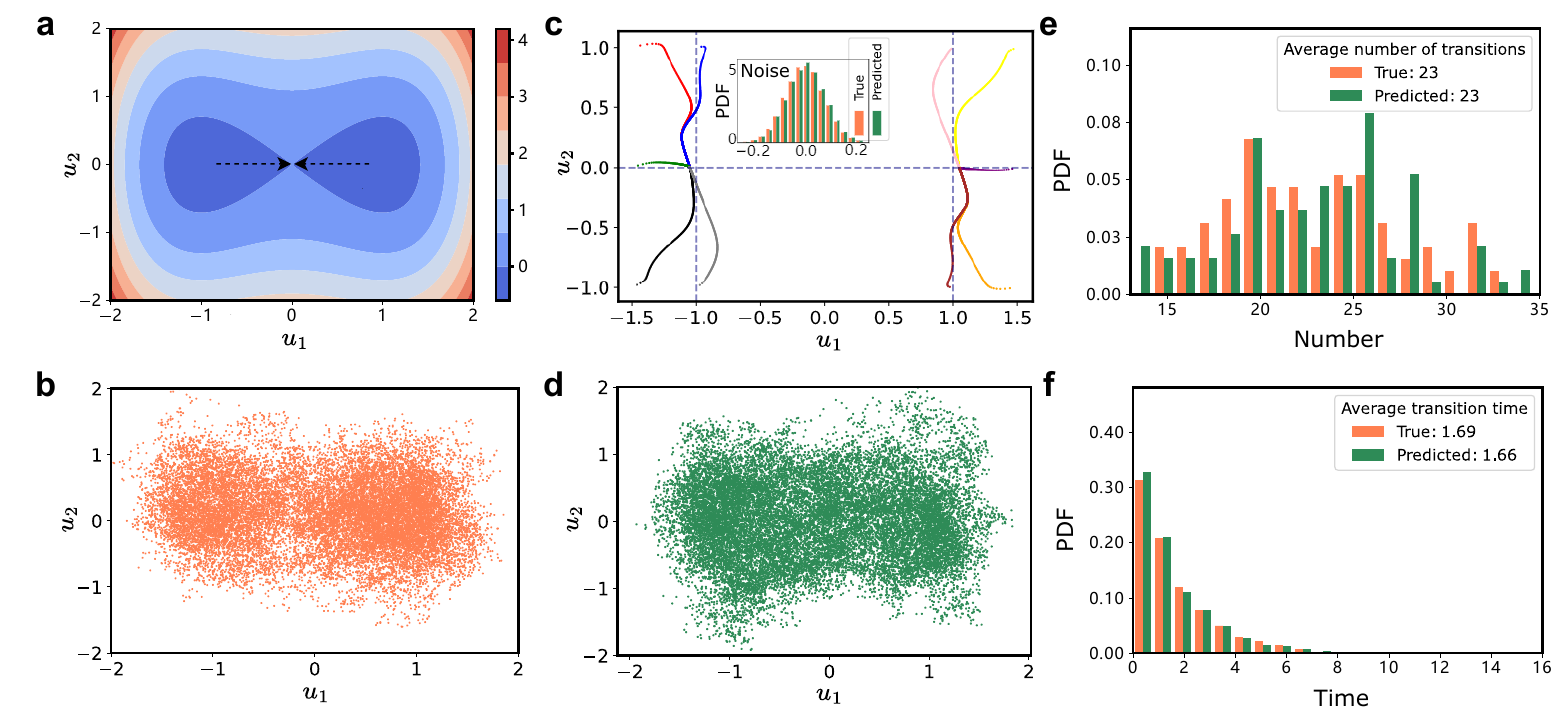}
    \caption{ \textbf{Capturing stochastic transitions in a 2D bistable gradient system with white noise.} The system is the same as Example 3 of the main text, and the parameter $a=0$. (a) Schematic of noise-induced transitions in the 2D bistable gradient system with Gaussian white noise. (b) Generated time series from Eqs.~(14) and (15) of the main text ($a=0, b=5, c=0,\varepsilon_1=\varepsilon_2=0.3, \delta t =0.002$) with $t=80$ as ground truth. (c) The trained slow-scale model transforms ten different start points into ten different slowly time-scale series (color lines), $t\in [40,80]$. Separate the noise distribution in the training phase. (d) Result of  prediction in $t\in [40,80]$ using the slow-scale model and noise in (c). (e) The number of transitions for the 100 replicates simulated in $t\in[40,80]$ and the 100 generated matches. The transition refers to the time series in the $u_1$-direction that crosses the zero point and remains either non-negative or non-positive for $50\delta t$. (f) Histograms of transition time for the test and predicted data. When a transition occurs within the system, the transition time is defined as the interval between two consecutive zero crossings in the $u_{1}$ direction.}
    \label{2d doublewell a=0}
\end{figure}

\clearpage
\begin{figure}[!htbp]
    \centering
\includegraphics[width=\textwidth]{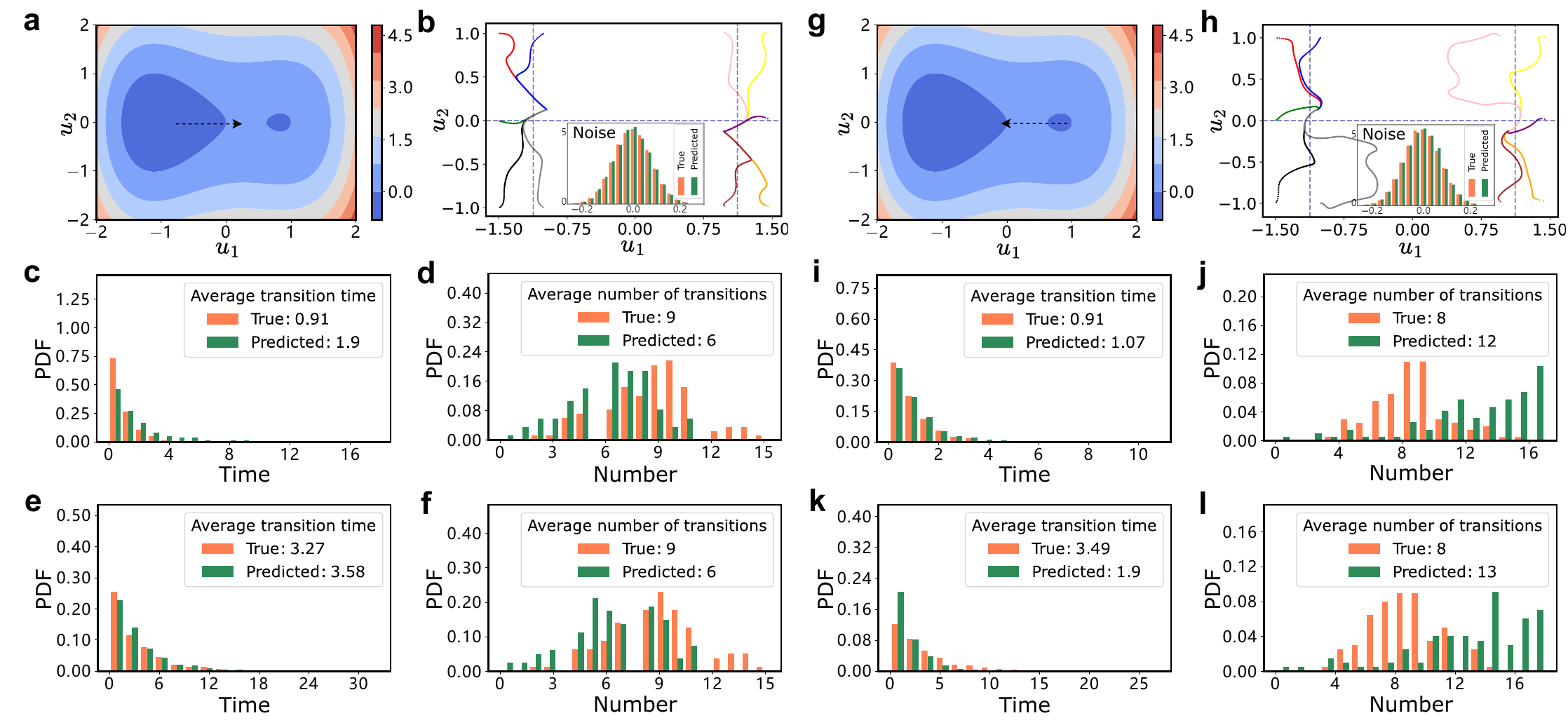} 
\caption{\textbf{Capturing stochastic transitions in a 2D tilted bistable gradient system with white noise.} The system is the same as Example 3 of the main text, and the parameter $c = 0.25$ making the potential tilted, $a = 0$, $\delta t = 0.002$. (a) Schematic of noise-induced upward transitions in the 2D tilted bistable gradient system. For figures (b-f), we focus on the upward transitions. (b) The trained slow-scale model transforms ten different start points into ten different slowly time-scale series (color lines), $t \in [40,80]$. Separate the noise distribution in the training phase. (c) Histograms of downward transition time for the test and predicted data. (d) The number of downward transitions for the test and predicted data. The duration of the predicting set is $20000\delta t$ (20000 data points). (e) Histograms of upward transition time for the test and predicted data. (f) The number of upward transitions for the test and predicted data. (g-l) Same as (a-f), and focusing on the downward transitions.}
    \label{2d doublewell with c}
\end{figure}

\clearpage
\begin{figure}[!htbp]
    \centering
    \includegraphics[width=\textwidth]{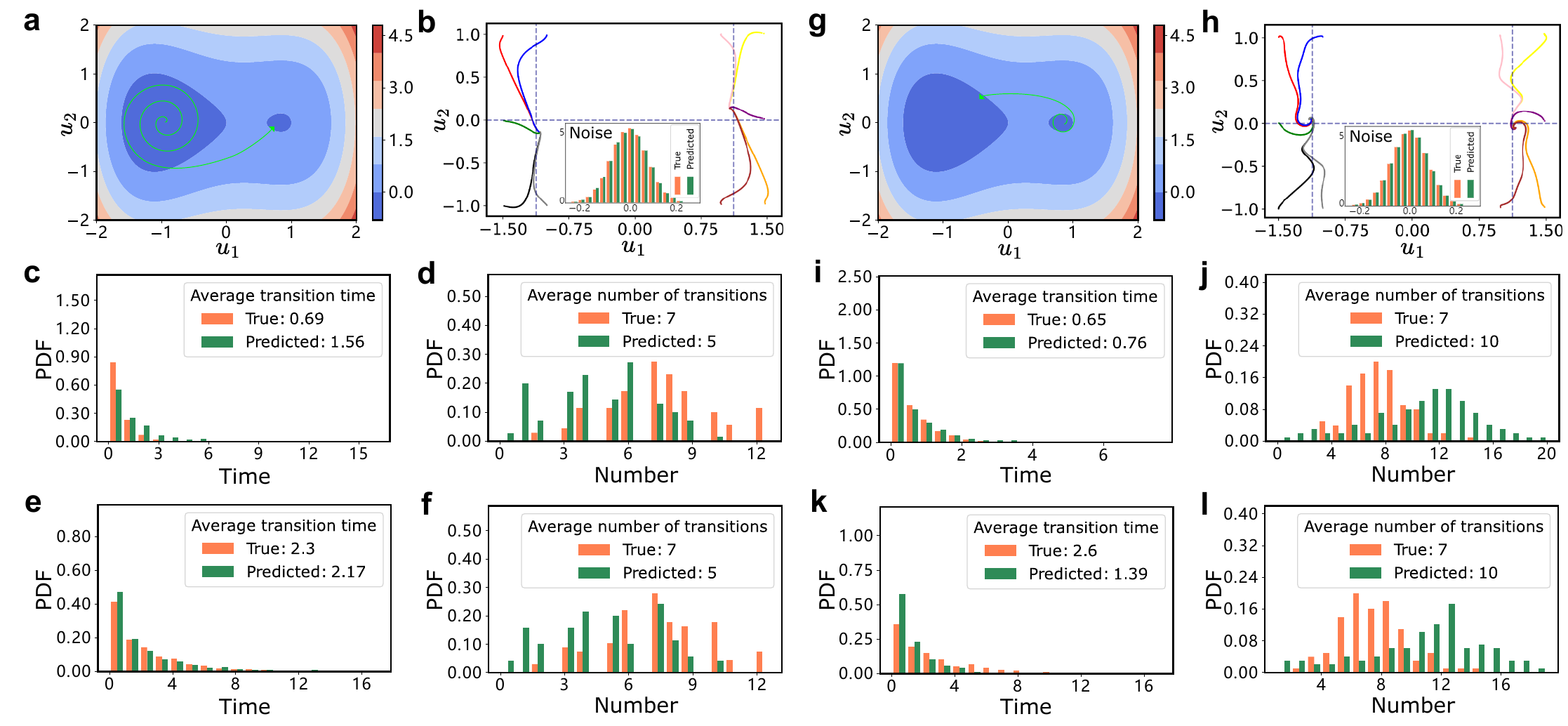}
  \caption{ \textbf{Learning noise-induced transitions in a 2D tilted bistable non-gradient system.} The system is the same as Example 3 of the main text, and the parameter $c = 0.25$ making the potential tilted, $a=2$ represents the strength of the non-detailed balance part, $\delta t = 0.001$. (a) Schematic of noise-induced upward transitions in the 2D tilted bistable non-gradient system with Gaussian white noise. For figures (b-f), we focus on the upward transitions. (b) The trained slow-scale model transforms ten different start points into ten different slowly time-scale series (color lines), $t \in [25,50]$. Separate the noise distribution in the training phase. (c) Histograms of downward transition time for the test and predicted data. (d) The number of downward transitions for the test and predicted data. The duration of the predicting set is $25000\delta t$ (25000 data points).  (e) Histograms of upward transition time for the test and predicted data. (f) The number of upward transitions for the test and predicted data. (g-l) Results focusing on the downward transitions. }
    \label{2d doublewell with a+c}
\end{figure}

\clearpage
\begin{figure}[!htbp]
    \centering
    \includegraphics[width=0.7\textwidth]{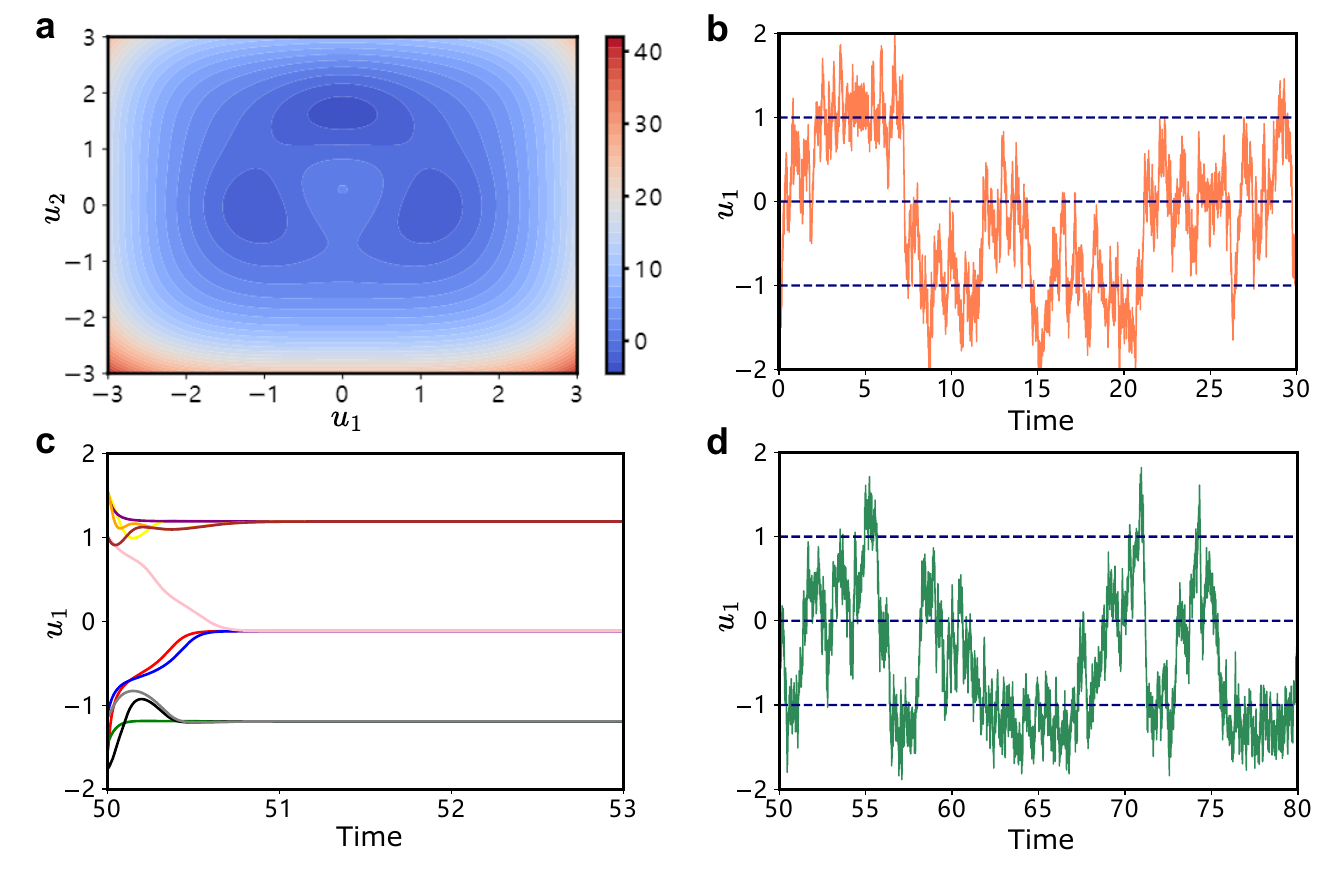}
    \caption{\textbf{Capturing stochastic transitions in a 2D tristable gradient system with white noise.} (a) Schematic of the 2D tristable system. (b) Generated time series from the \cref{2-D triple in x,2-D triple in y} ($\gamma=1,  \delta t=0.001$) of \cite{belkacemi2021chasing} with $t=30$ as ground truth. In the scenario where $\gamma=1$, the three potential wells exhibit comparable depths. (c) The trained slow-scale model transforms ten different start points into ten different slowly time-scale series (color lines), only the $u_{1}$ direction is depicted. (d) Result of  prediction using the slow-scale model and separated noise in $t\in[50,80]$.
    }
    \label{2D-triple fig}
\end{figure}

\clearpage
\section{Supplementary tables}

\begin{table}[!htbp]
  \centering
  \caption{Hyperparameters for \autoref{Se1}. }
  \begin{tabular}{@{\extracolsep{4pt}}lllllllllll@{\extracolsep{4pt}}}
    \toprule
    \toprule
  Set & Transitions & $\delta t$ & $T_{\text{train}}$ (Time steps)&$T_{\text{predict}}$ (Time steps) &$N$&$K_{\text{in}}$&$D$& $\rho$& $\alpha$&$\beta$\\
    \midrule
   1& Upward & 0.01 & 10000 & 10000 &800 &4.2 &4 & $1.1 \times 10^{-3}$&0.18&$1 \times 10^{-8}$ \\
    2&Downward & 0.01 & 10000 & 10000 &800 &3.8 &4 & $1.2 \times 10^{-3}$ & 0.25&$1 \times 10^{-8}$ \\
   
    \bottomrule
    \bottomrule
  \end{tabular}
  \label{doublewell tilted hyperparameters}
\end{table}

\begin{table}[!htbp]
  \centering
  \caption{Hyperparameters for \autoref{Se2}.}
  \begin{tabular}{@{\extracolsep{4pt}}lllllllll@{\extracolsep{4pt}}}
    \toprule
    \toprule
     $\delta t$ & $T_{\text{train}}$ (Time steps)&$T_{\text{predict}}$ (Time steps) &$N$&$K_{\text{in}}$&$D$& $\rho$& $\alpha$&$\beta$\\
    \midrule
     0.002 & 20000 & 20000 &1200 &1.5 &2 & $1.6 \times 10^{-3}$ &0.36&$1 \times 10^{-7}$ \\   
    \bottomrule
    \bottomrule
  \end{tabular}
  \label{2d doublewell a0 hyperparameters}
\end{table}

\begin{table}[!htbp]
  \centering
  \caption{Hyperparameters for \autoref{Se3} and \autoref{Se4}.}
  \begin{tabular}{@{\extracolsep{4pt}}lllllllllll@{\extracolsep{4pt}}}
    \toprule
    \toprule
    Set & Transitions & $\delta t$ & $T_{\text{train}}$ (Time steps)&$T_{\text{predict}}$ (Time steps)  &$N$&$K_{\text{in}}$&$D$& $\rho$& $\alpha$&$\beta$\\
    \midrule
  1&  Upward & 0.002 & 20000 & 20000 &1200 &1.6 &3.6 & $1.4 \times 10^{-3}$&0.58&$1 \times 10^{-7}$ \\
 2&  Downward & 0.002 & 20000 & 20000 &1200 &1.5 &3.5 & $1.1 \times 10^{-3}$ & 0.64&$1 \times 
    10^{-7}$ \\
   \midrule
   3& Upward & 0.001 & 25000 & 25000 &1200 &1.1 &3 & $9.1 \times 10^{-4}$ &0.52&$1 \times 10^{-7}$ \\
   4& Downward & 0.001 & 25000 & 25000&1200 &1.3 &3 & $8.9 \times 10^{-4}$ & 0.6&$1 \times 
    10^{-7}$ \\
   
    \bottomrule
    \bottomrule
  \end{tabular}
  \label{2d doublewell tilted  hyperparameters}
\end{table}

\begin{table}[!htbp]
  \centering
  \caption{Hyperparameters for \autoref{Se5}.}
  \begin{tabular}{@{\extracolsep{4pt}}lllllllll@{\extracolsep{4pt}}}
    \toprule
    \toprule
     $\delta t$ & $T_{\text{train}}$ (Time steps)&$T_{\text{predict}}$ (Time steps)  &$N$&$K_{\text{in}}$&$D$& $\rho$& $\alpha$&$\beta$\\
    \midrule
  0.001 & 50000 & 50000 &1200 &1 &2.8 & $1.7 \times 10^{-3}$&0.45 &$1 \times 10^{-6}$\\
    \bottomrule
    \bottomrule
  \end{tabular}
  \label{Triple Hyperparameters}
\end{table}

\end{document}